\documentclass[fleqn,usenatbib]{mnras}

\usepackage{newtxtext,newtxmath}
\usepackage[T1]{fontenc}

\DeclareRobustCommand{\VAN}[3]{#2}
\let\VANthebibliography\thebibliography
\def\thebibliography{\DeclareRobustCommand{\VAN}[3]{##3}\VANthebibliography}

\usepackage{graphicx}	% Including figure files
\usepackage{amsmath}	% Advanced maths commands
\usepackage{xcolor}

\title[BU CMi - a tight quadruple system]{BU Canis Minoris -- the Most Compact Known Flat Doubly Eclipsing Quadruple System}

\author[T. Pribulla et al.]{
Theodor Pribulla,$^{1}$\thanks{E-mail: pribulla@ta3.sk}
Tam\'as Borkovits,$^{2,3,4,5,6}$
Rahul Jayaraman,$^{7}$
Saul Rappaport,$^{7}$
Tibor Mitnyan,$^{2,3,8}$
\newauthor
Petr Zasche,$^{9}$
Richard Kom\v{z}{\'i}k,$^{1}$
Andr\'as P\'al,$^{10}$
Robert Uhla\v{r},$^{11}$
Martin Ma\v{s}ek,$^{11,12}$
Zbyn\v{e}k Henzl,$^{11,13}$
\newauthor
Imre Barna B\'{\i}r\'o,$^{2,3}$
Istv\'an Cs\'anyi,$^{2}$
Remko Stuik,$^{14}$
Martti H.\,Kristiansen,$^{15}$
Hans M.\,Schwengeler,$^{16}$
\newauthor
Robert Gagliano,$^{17}$
Thomas L.\,Jacobs,$^{18}$
Mark Omohundro,$^{16}$
Veselin Kostov,$^{19,21}$
Brian P. Powell,$^{19}$
\newauthor
Ivan A. Terentev,$^{20}$
Andrew Vanderburg,$^{7}$
Daryll LaCourse,$^{22}$
Joseph E.\,Rodriguez,$^{23}$
G\'asp\'ar Bakos,$^{24}$
\newauthor
Zolt\'an Csubry,$^{24}$
Joel Hartman$^{24}$
\\
$^{1}$Astronomical Institute of the Slovak Academy of Sciences, 059 60 Tatranská Lomnica, Slovakia\\
$^{2}$Baja Astronomical Observatory of Szeged University, H-6500 Baja, Szegedi \'ut, Kt. 766, Hungary\\
$^{3}$ELKH-SZTE Stellar Astrophysics Research Group, H-6500 Baja, Szegedi \'ut, Kt. 766, Hungary\\
$^{4}$Konkoly Observatory, Research Centre for Astronomy and Earth Sciences, H-1121 Budapest, Konkoly Thege Miklós \'ut 15-17, Hungary\\
$^{5}$ELTE Gothard Astrophysical Observatory, H-9700 Szombathely, Szent Imre h. u. 112, Hungary\\
$^{6}$MTA-ELTE Exoplanet Research Group, H-9700 Szombathely, Szent Imre h. u. 112, Hungary\\
$^{7}$MIT Kavli Institute and Department of Physics, 77 Massachusetts Avenue, Cambridge, MA 02139, USA\\
$^{8}$Department of Experimental Physics, University of Szeged, 6720 Szeged, D\'om t\'er 9, Hungary\\
$^{9}$Charles University, Faculty of Mathematics and Physics, Astronomical Institute, V Hole\v{s}ovi\v{c}k\'ach 2, Praha 8, 180 00, Czech Republic\\
$^{10}$Konkoly Observatory, Research Centre for Astronomy and Earth Sciences, MTA Centre of Excellence, Konkoly Thege Mikl\'os  \'ut 15-17, \\ H-1121 Budapest, Hungary\\
$^{11}$Variable Star and Exoplanet Section, Czech Astronomical Society, Fri\v{c}ova 298, 251 65 Ond\v{r}ejov, Czech Republic\\
$^{12}$FZU - Institute of Physics of the Czech Academy of Sciences, Na Slovance 1999/2, CZ-182 00, Praha, Czech Republic\\
$^{13}$Hv\v{e}zd\'arna Jaroslava Trnky ve Slan\'em, Nosa\v{c}ick\'a 1713, Slan\'y 1, 274 01, Czech Republic\\
$^{14}$Leiden Observatory, Leiden University, Postbus 9513, 2300 RA Leiden, The Netherlands\\
$^{15}$Brorfelde Observatory, Observator Gyldenkernes Vej 7, DK-4340 T\o ll\o se, Denmark\\
$^{16}$Citizen Scientist, c/o Zooniverse, Dept,~of Physics, University of Oxford, Denys Wilkinson Build., Keble Road, Oxford, OX1 3RH, UK\\
$^{17}$Amateur Astronomer, Glendale, AZ 85308, USA\\
$^{18}$Amateur Astronomer, 12812 SE 69th Place Bellevue, WA 98006, USA\\
$^{19}$NASA Goddard Space Flight Center, 8800 Greenbelt Road, Greenbelt, MD 20771, USA\\
$^{20}$Citizen Scientist, Planet Hunter, Petrozavodsk, Russia\\
$^{21}$SETI Institute, 189 Bernardo Avenue, Suite 200, Mountain View, CA 94043, USA\\
$^{22}$Amateur Astronomer, 7507 52nd Place NE Marysville, WA 98270, USA\\
$^{23}$Center for Data Intensive and Time Domain Astronomy, Department of Physics and Astronomy, Michigan State University, East Lansing, MI 48824, USA\\
$^{24}$133 Peyton Hall, Astrophysical Sciences, Princeton University, 4 Ivy Ln, Princeton, NJ 08544
}

\date{Accepted XXX. Received YYY; in original form ZZZ}

\pubyear{2023}

\begin{document}
\label{firstpage}
\pagerange{\pageref{firstpage}--\pageref{lastpage}}
\maketitle

\begin{abstract}
We have found that the 2+2 quadruple star system BU CMi is currently the most compact quadruple system known, with an extremely short outer period of only 121 days.  The previous record holder was TIC 219006972 \citep{kostovetal23}, with a period of 168 days.  The quadruple nature of BU CMi was established by \citet{2021ARep...65..826V}, but they misidentified the outer period as 6.6 years.  BU CMi contains two eclipsing binaries (EBs), each with a period near 3 days, and a substantial eccentricity of $\simeq 0.22$.  All four stars are within $\sim$0.1 M$_\odot$ of 2.4 M$_\odot$.  Both binaries exhibit dynamically driven apsidal motion with fairly short apsidal periods of $\simeq 30$ years, thanks to the short outer orbital period.  The outer period of 121 days is found both from the dynamical perturbations, with this period imprinted on the eclipse timing variations (ETV) curve of each EB by the other binary, and by modeling the complex line profiles in a collection of spectra.  We find that the three orbital planes are all mutually aligned to within 1 degree, but the overall system has an inclination angle near $83.5^\circ$.  We utilize a complex spectro-photodynamical analysis to compute and tabulate all the interesting stellar and orbital parameters of the system.  Finally, we also find an unexpected dynamical perturbation on a timescale of several years whose origin we explore.  This latter effect was misinterpreted by \citet{2021ARep...65..826V} and led them to conclude that the outer period was 6.6 years rather than the 121 days that we establish here. 
\end{abstract}


%%%%%%%%%%%%%%%%%%%%%%%%%%%%%%%%%%%%%%%%%%%%%%%%%%

%%%%%%%%%%%%%%%%% BODY OF PAPER %%%%%%%%%%%%%%%%%%

\section{Introduction} 
\label{sec:intro}

There are currently more than 300 known 2+2 quadruples consisting of an orbiting pair of eclipsing binaries, most of which have been found with \textit{TESS} (\citealt{2022ApJS..259...66K,zasche22,kostovetal23}. The criteria for accepting these as quadruples are: (i) there are two eclipsing binaries that are (ii) unresolved at the pixel level with \textit{TESS}, and (iii) which show only one dominant star in Gaia within the \textit{TESS} pixel.  However, given that Gaia does not often distinguish stars that are $\lesssim 1/2''$ apart, and that these objects are typically a kpc away, this implies only that the physical separation of the EBs is $\lesssim 500$ AU or so.  The corresponding outer orbital periods are only constrained to an order of $\lesssim 5000$ years.

At the largest of these orbital separations, the quadruples would be too wide for easy-to-measure dynamical interactions that could lead to a determination of the outer orbital period.  However, a small percentage of these quadruples have much closer separations of less than a few AU, and these have led to measurable outer orbits as well as other interesting dynamical interactions, such as TIC 454140642 (432 days; \citealt{kostovetal21}), TIC 219006972 (168 days; \citealt{kostovetal23}), and  VW~LMi (355 days; \citealt{VWLMi-old}).

With quadruples having outer orbital periods as short as $\lesssim$ a couple of years, interesting and informative dynamical interactions to look for include: (i) dynamical delays resulting from a changing period of either EB due to the varying distance to the other EB; (ii) dynamically forced orbital precession in the EBs; and (iii) forced precession of the orbital planes leading to eclipse depth variations.

Every once in a while, one of these quadruples turns out to have a dramatically short outer period (e.g. TIC 219006972; \citealt{kostovetal23}) and it becomes quite feasible to use ETV data as well as RV data to completely diagnose most of the important stellar and orbital parameters. In this work, we report on BU CMi, a 2+2 quadruple which we find to have the shortest known outer orbital period of 121 days, and a very interesting array of dynamical effects.

This work is organized as follows. In the remainder of Sect.~\ref{sec:intro}, we discuss how the quadruple nature of BU CMi came to be known. Section ~\ref{sec:observations} briefly reviews the observations that we bring to bear on the analysis of BU CMi, including with \textit{TESS} and follow-up ground-based photometry and spectroscopy. In Sect.~\ref{sec:RVs} we briefly describe our use of broadening functions to extract radial velocities (RVs) from the spectra. Section \ref{sec:Period_study} details our production of a long-term ETV curve from archival as well as \textit{TESS} data, and what we can learn from a visual inspection of it. In Sect.~\ref{sec:Theo_analyses} we explain in some detail how we fit a model of stellar and orbital parameters directly to the complex and overlapping line profiles in each spectrum.  Section \ref{sec:spectral_disentanglement} is devoted to a discussion of disentangling the four spectra, after we determined the stellar and orbital parameters.  We review our spectro-photodynamical model for ascertaining the system parameters in Sect.~\ref{sect:photodynamics}. A discussion of our work and its implications follows in Sect.~\ref{sec:discussion}.  We give some concluding remarks in Section \ref{sec:conclusions}.

\subsection{Prior work on Multiple Stars}
The Transiting Exoplanet Survey Satellite (\textit{TESS}; \citealt{2015JATIS...1a4003R}) has been instrumental in the discovery of multiple star systems. There has been significant progress on the problem of identifying triple, quadruple, quintuple, and even sextuple star systems from \textit{TESS} data; for instance, the Planet Hunters citizen science project uncovered tens of multiple star system candidates \citep{2021MNRAS.501.4669E}. In parallel, a combination of machine learning techniques and human vetting led to the creation of an extensive catalog of quadruple stellar systems %\footnote{\ron{What about Andrei's MSC and Petr's former cataolg? This must be reworded!!!}} 
\citep{2022ApJS..259...66K}, in addition to the identification of the first-ever sextuply-eclipsing six star system \citep{2021AJ....161..162P}.

\subsection{BU Canis Minoris}

BU~CMi (HD 65241, HIP 38945, TIC 271204362) is listed as a suspected Algol-type eclipsing binary in the 74th special namelist of new variable stars \citep{74namelist}. In spite of its brightness ($V$ = 6.42), the object was not subject to any detailed study until \citet{2021ARep...65..826V} found it to be a quadruple system composed of two eclipsing binaries with $P_A$ = 2.94 days and $P_B$ = 3.26 days. Although the authors had observed BU~CMi with a photoelectric photometer in 2012, they first noticed two systems of eclipses in  \textit{MASCARA} photometric data in 2020. According to their analysis, all four stars in the quadruple are similar (A0 spectral type), having masses between 3.1 - 3.4 M$_\odot$.  The orbital eccentricities of the inner binaries were found to be relatively high for close binaries, with $e_A$ = 0.20 and $e_B = 0.22$.  The authors interpreted the observed ETVs as being due to (i) relatively rapid apsidal motion in the inner (i.e., binary) orbits with the periods $U_A$ = 25.4 years and $U_B$ = 26.3 years, and (ii) light travel time effects (LTTE) from the outer orbit between the two binaries having a period of 6.62-years and a high-eccentricity ($e_{AB}$ = 0.7). 

Gaia DR3 \citep{Gaia2022} lists BU~CMi as a single object astrometrically and spectrosopically with a parallax of $\pi$ = 4.0143$\pm$0.0335 mas and the following atmospheric parameters $T_{\rm eff} = 10173^{+43}_{-39}$ K, $\log g = 3.727^{+7}_{-6}$, and [M/H] = 0.778$^{+17}_{-40}$.

The main catalogued photometric and kinematic data for BU CMi are given in Table~\ref{tab:EBparameters}.

\subsection{Our two independent discoveries of the quadruple nature of BU CMi}

The Visual Survey Group (`VSG'), in its search for compact multistellar systems, continues to visually survey large numbers of light curves from the \textit{TESS} mission \citep{kristiansen22}.  Some of its findings, including in the area of multistellar systems, are given in \citet{kristiansen22}, and \citet{rappaport19,rappaport22}.  In 2021 March, the group spotted BU CMi in the Sector 34 light curves and immediately identified it as a potential quadruple star system.

In parallel to the VSG's survey of the light curves from the \textit{TESS} full-frame images (FFIs), RJ and SR have also been searching the 2-minute and 20-second cadence light curves for strongly periodic or time-varying stellar phenomena. Every 2-minute-cadence light curve is passed through an algorithm that flags it for further review if there is a detection of at least a 12-$\sigma$ peak in its periodogram. When searching the Sector 34 short-cadence light curves through this algorithmic process, BU~CMi triggered on this algorithm; upon human review, this light curve was found to be a bona fide quadruple system and flagged for further follow-up.

Since that time, we have continued to collect information on, and model, the BU CMi system. After two years of study, we now report here on the discovery of a 121 day outer period for the system.

\begin{table}
\caption{Main catalog data for BU CMi\label{tab:EBparameters}}
\begin{tabular}{lrrc}
Parameter & Value & Error & Source \\
\\
\hline
\multicolumn{4}{l}{\bf Identifying Information} \\
\hline
TIC ID & 271204362 & & 1 \\
Gaia ID & 3144498015858945408 & & 2 \\
$\alpha$  (J2000, hh:mm:ss) & 07:58:05.887 &  & 2 \\
$\delta$  (J2000, dd:mm:ss) & +07:12:48.51 &  & 2 \\
$\mu_{\alpha}$ (mas~yr$^{-1}$) & $-9.42671$ & 0.03876 & 2 \\
$\mu_{\delta}$ (mas~yr$^{-1}$) & $-11.60272$ & 0.02780 & 2 \\
$\varpi$ (mas) & 4.01432 & 0.03353 & 2 \\
Distance (pc) & 247 & 2 & 3\\
RUWE & 0.84464 & & 2 \\
E(B-V) & 0.00733 & 0.00138 & 1 \\
$T_\mathrm{eff}$ (K)  & 10200 & 178 & 1 \\
& 10175 & 41 & 2 \\
\hline
\multicolumn{4}{l}{\bf Photometric Properties} \\
\hline
$T$ (mag) & 6.4556 & 0.0082 & 1 \\
$B$ (mag) & 6.411 & 0.024 & 1\\
$V$ (mag) & 6.417 & 0.023 & 1 \\
$B_T$ (mag) & 6.410 & 0.10 & 4\\
$V_T$ (mag) & 6.417 & 0.010 & 4\\
$Gaia$ (mag) & 6.4046 &  0.0015 & 2 \\
$G_{BP}$ (mag) & 6.4076 & 0.0031 & 2 \\
$G_{RP}$ (mag) & 6.4347 & 0.0078 & 2 \\
$g'$ (mag) & 6.41 & 0.03 & 5\\
$r'$ (mag) & 6.63 & 0.03 & 5\\
$i'$ (mag) & 6.80 & 0.03 & 5\\
$z'$ (mag) & 6.91 & 0.03 & 5\\
$J$ (mag) & 6.433 & 0.019 & 6 \\
$H$ (mag) & 6.433 & 0.026 & 6 \\
$K$ (mag) & 6.466 & 0.020 & 6 \\
$W1$ (mag) & 6.467 & 0.023 & 7 \\
$W2$ (mag) & 6.420 & 0.023 & 7 \\
$W3$ (mag) & 6.536 & 0.015 & 7 \\
$W4$ (mag) & 6.490 & 0.067 & 7 \\
\hline
\end{tabular}
Sources: (1) TIC-8 \citep{TIC}, (2) Gaia DR3 \citep{Gaia2022}, (3) \citet{2021AJ....161..147B} (4) Tycho-2 catalog \citep{Tycho2}, (5) \citet{2008PASP..120.1128O} (6) 2MASS All-Sky Catalog of Point Sources \citep{2MASS}, (7) AllWISE catalog \citep{WISE}
\end{table}

\section{Observations}
\label{sec:observations}
\subsection{\textit{TESS} Observations}
BU CMi (TIC\,271204362) was observed during \textit{TESS} Sectors 7, 34, and 61 (from 7 January 2019 to 2 February 2019, 13 January 2021 to 9 February 2021, and 18 January 2023 to 12 February 2023, respectively). In Sector 7, it was observed at 30-minute cadence in the full-frame images (FFIs); while in Sectors 34 and 61, besides the 600 and 200-sec cadence FFIs, it was also observed at 2-minute cadence. For most of our study we used the FFI observations, from which the light curves were processed using the {\sc FITSH} pipeline \citep{FITSH12}. We used the 2-min cadence SAP-FLUX light curves, which were downloaded directly from the Barbara A. Mikulski Archive for Space Telescopes (MAST) website, only for the determination of mid-eclipse times over Sectors 34 and 61 data. Naturally, in the case of the Sector 7 observations, we could determine the mid-eclipse times only from the sparsely cadenced FFI data at 30 min.

\begin{figure}
    \centering
    \includegraphics[width=1.0\linewidth]{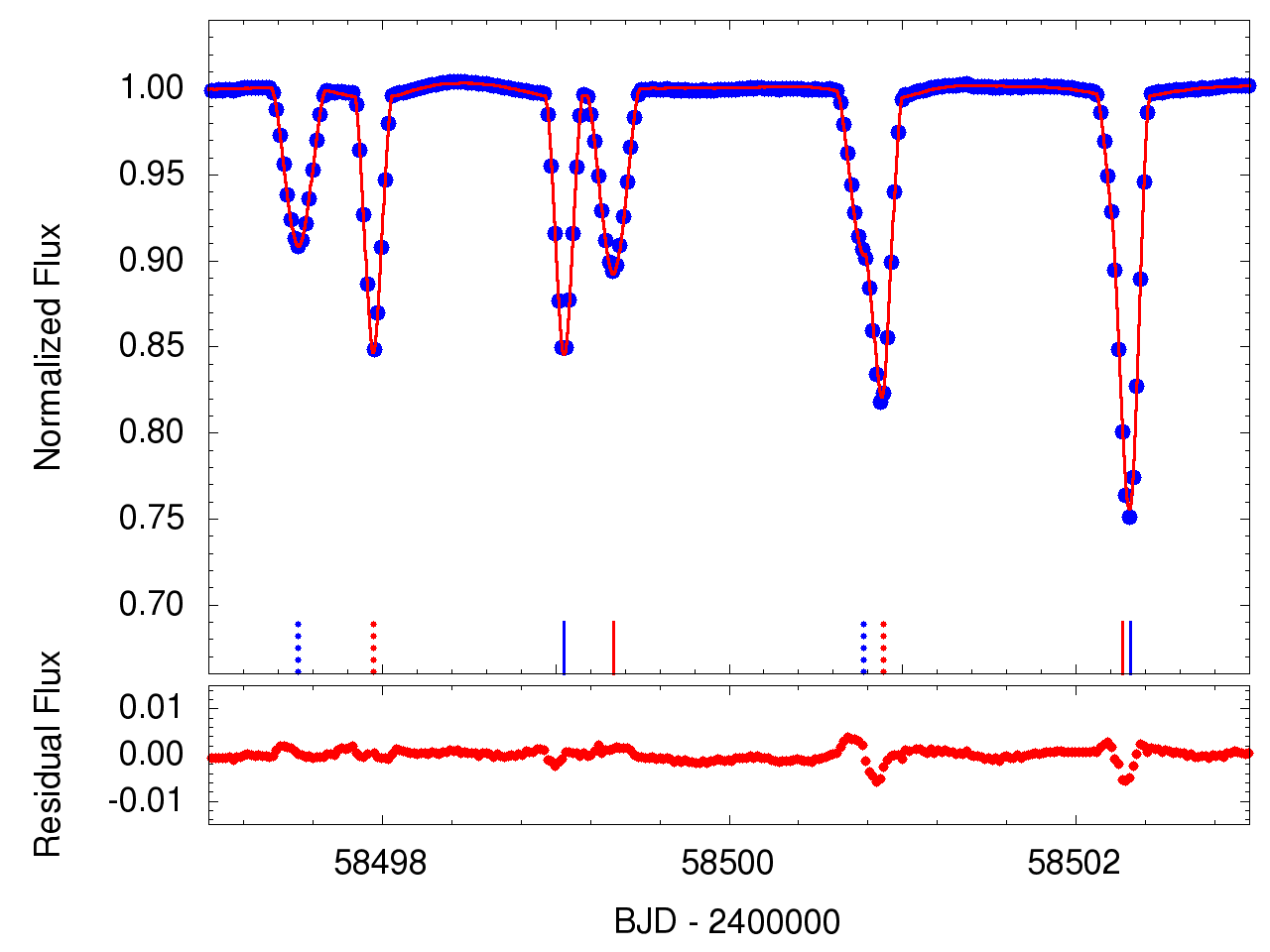}
    \includegraphics[width=1.0\linewidth]{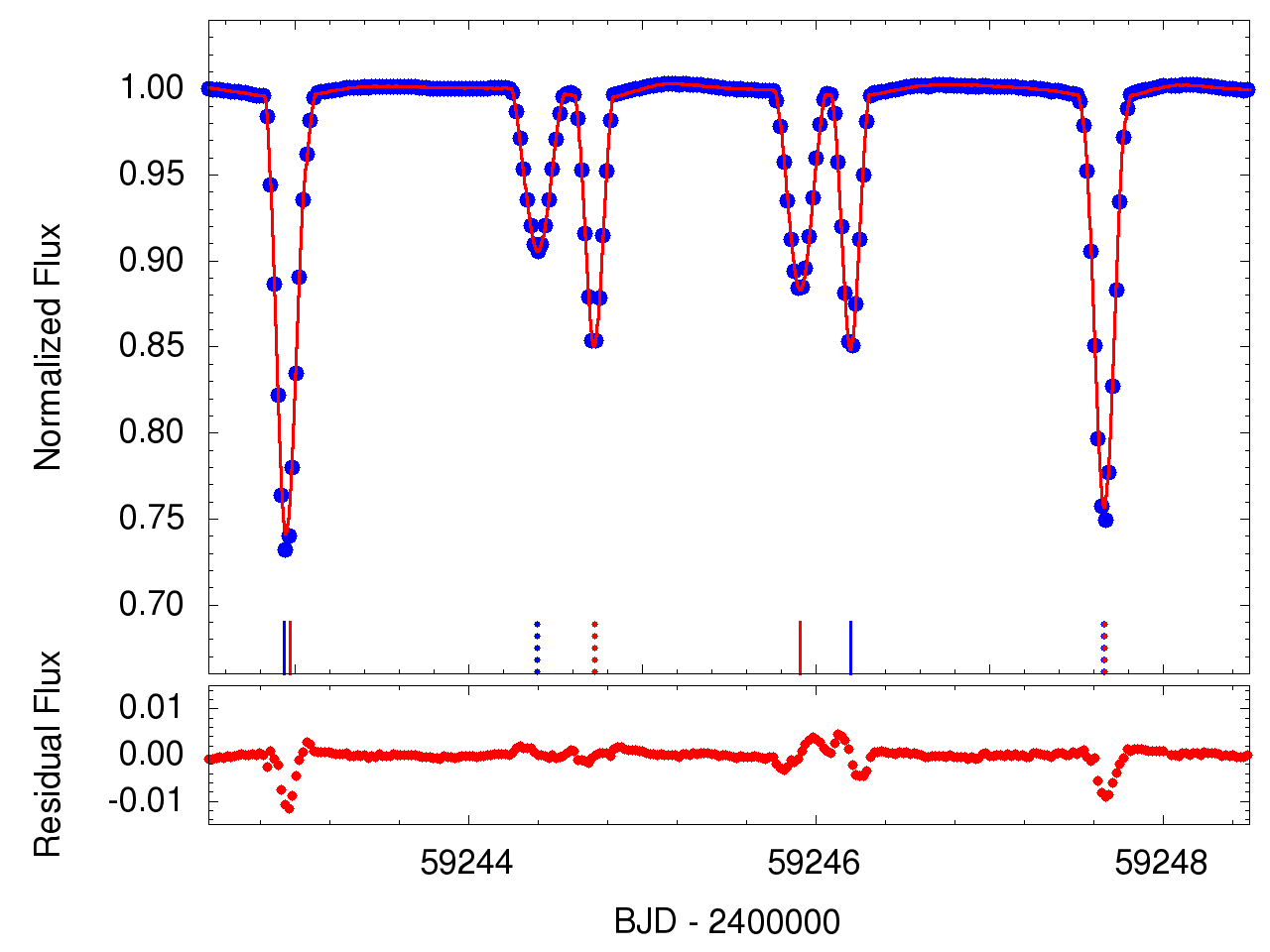}
    \includegraphics[width=1.0\linewidth]{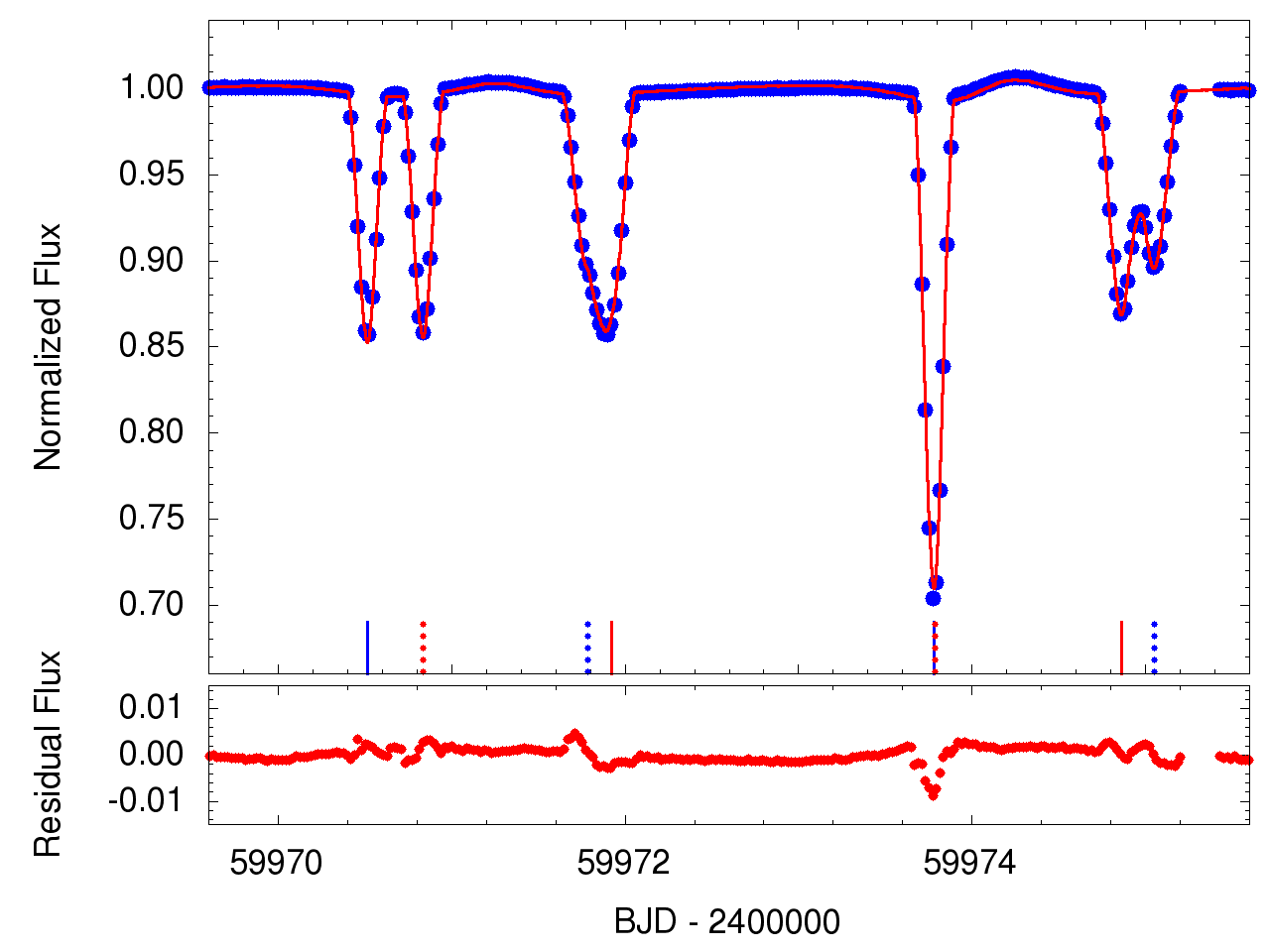}
   \caption{Segments of \textit{TESS} data from sectors 7, 34 and 61 (top, middle and bottom, respectively). The superposed model is shown with a red curve (see Sect.~\ref{sect:photodynamics}). Solid and dashed vertical lines show the primary and secondary mid-eclipses of binaries A (red) and B (blue). In each panel, residuals of the fit are plotted at the bottom.}
   \label{fig:TESSlightcurves+model}
\end{figure}

We show some illustrative segments of the three \textit{TESS} sectors in Fig.~\ref{fig:TESSlightcurves+model}. The superposed model curves will be discussed in Sect.~\ref{sect:photodynamics}.

 % Fig. 1

\subsection{Ground-based Photometry}
BU CMi is bright enough ($M_V$ = 6.42) that it can be reliably observed using small-aperture telescopes from the ground. We used archival data obtained from the Kilodegree Extremely Little Telescope (\textit{KELT}--2600 points; \citealt{2007PASP..119..923P, 2012PASP..124..230P}), the Multi-site All-Sky CAmeRA (\textit{MASCARA}--10800 points; see \citealt{2017A&A...601A..11T} and \citealt{2018A&A...617A..32B}), and the Hungarian Automated Telescope Network (\textit{HATnet}--3200 points; \citealt{2004PASP..116..266B}). BU CMi was heavily saturated in the \textit{HATNet} observations, so we performed a custom analysis of these data that differs from the standard methods used by the survey. We extracted aperture photometry for BU CMi and 200 other comparably bright stars through an annulus excluding the saturated core of the target. We then performed an ensemble correction using a linear combination of the 200 bright neighbors to correct for instrumental and atmospheric variations in the resulting light curve.

Additionally, we carried out further photometric follow-up observations between 2021 March and 2023 April. These were mainly obtained with a small 34-mm telescope in a private observatory in J\'{\i}lov\'e u Prahy in the Czech Republic, as well as remotely in Northern Italy (both by one of the co-authors R.U.). Another observing site was in Argentina as part of the Pierre Auger Observatory \citep{2021JInst..16P6027A}, and obtained by M.M. On three additional nights, the target was also observed by Z.H. from two observing sites in the Czech Republic. Finally, two further eclipses of BU CMi were observed with the RC80 telescope of Baja Astronomical Observatory in 2022 and 2023.

\subsection{Spectroscopy from Skalnat\'e Pleso Observatory}

High-dispersion spectroscopy was obtained with a 1.3 m, f/8.36 Nasmyth-Cassegrain telescope equipped with a fiber-fed échelle spectrograph at the Skalnat\'e Pleso (SP) Observatory, Slovakia. Its layout follows the MUSICOS design \citep{1992A&A...259..711B}. The spectra were recorded by an Andor iKon-L DZ936N-BV CCD camera with a 2048 $\times$ 2048 array, 13.5 $\mu$m square pixels, 2.9\,e$^-$ readout noise, and a gain close to unity. The spectral range of the instrument is 4250–7375~\AA~ (56 \'echelle orders), with a maximum resolution of $R$ = 38\,000. Because of the relatively long orbital period of both inner binaries, three 600-sec exposures were combined to increase the SNR and to clean cosmic ray hits. The raw spectroscopic data were reduced as in \citet{2015AN....336..682P} using IRAF package tasks, LINUX shell scripts, and FORTRAN programs. In the first step, master dark and flat-field frames were produced, based on the spectra from the tungsten lamp and blue LED. In the second step, the photometric calibration of the frames was performed using dark and flat-field frames. Bad pixels were cleaned using a bad-pixel mask, and cosmic ray hits were removed using the program of \citet{2004PASP..116..148P}. Order positions were defined by fitting sixth-order Chebyshev polynomials to tungsten-lamp and blue LED spectra. Subsequently, scattered light was modelled and subtracted, and then aperture spectra were extracted for both BU CMi and the ThAr lamp. The resulting two-dimensional spectra were then dispersion solved and combined to one-dimensional spectra.  Finally, all spectra were continuum normalized. The typical radial velocity stability of the spectrograph is 200 m\,s$^{-1}$.

In total, 60 spectra with per-pixel SNR ranging from 45 to 175 (at 5500~\AA) were obtained from 2020 March 17 through 2023 March 3.

\subsection{CTIO spectroscopy}

Additional spectroscopy was obtained with the CHIRON fiber-fed \'{e}chelle spectrograph at the 1.5m telescope of the Cerro-Tololo Interamerican Observatory, Chile.  A detailed description of the spectrograph can be found in \citet{chiron}. All spectra were taken in the slicer mode providing $R$ = 80\,000.  The spectrum was extracted from 70 \'{e}chelle orders covering 4080-7000\,\AA. Similar to the case of the SP spectroscopy, three consecutive 600-second exposures were combined. Even without an iodine cell, the spectrograph stability is a few m\,s$^{-1}$. Because of a reflection causing a blemish in the blue part of the spectrum, an independent pipeline based on the IRAF scripts has been developed to make use of the full spectral range of the instrument. In addition to the reduction steps taken for the SP spectra, the raw frames were trimmed and corrected for the overscan. The relative responses of four quadrants as read out by different amplifiers were taken into account. In total, BU~CMi was observed on 16 nights from 2022 November 11, until 2023 March 29. The per-pixel SNR ranged from 70 to 290 (at 5500~\AA).

\subsection{Konkoly and Rozhen spectroscopic observations}
\label{sec:KonkolyRozhen}

Additional spectroscopic measurements were obtained with the $R$\,$\sim$\,20\,000 échelle spectrograph mounted on the 1m RCC telescope at Konkoly Observatory, Hungary between 2021 December and 2022 March. The spectrograph is capable of covering the 3890-8670\,\AA\, wavelength range in a set of 33 échelle orders. The images were taken with a back-illuminated FLI ML1109 CCD camera having an array of $2048 \times 506$ pixels of size 12\,$\mu$m, 10 e$^-$ readout noise, and a gain close to unity. A total of 11 spectra were observed, with exposure times varying between 900 and 3600 seconds. Two additional spectra were taken with the 2\,m RCC telescope at NAO Rozhen, Bulgaria equipped with the $R$\,$\sim$\,30\,000 ESpeRo spectrograph in January 2022. The details of the instruments can be found in \citet{bonev17}. The spectra were reduced completely in the same manner as in \citet{mitnyan20} including the steps of bias, dark and flat corrections, wavelength calibration, continuum normalization, telluric line removal, and barycentric correction. The per-pixel SNR of these additional observed spectra is between 70 and 140.

\section{Radial velocities}
\label{sec:RVs}

\begin{figure}
    \centering
    \includegraphics[width=1.0\linewidth]{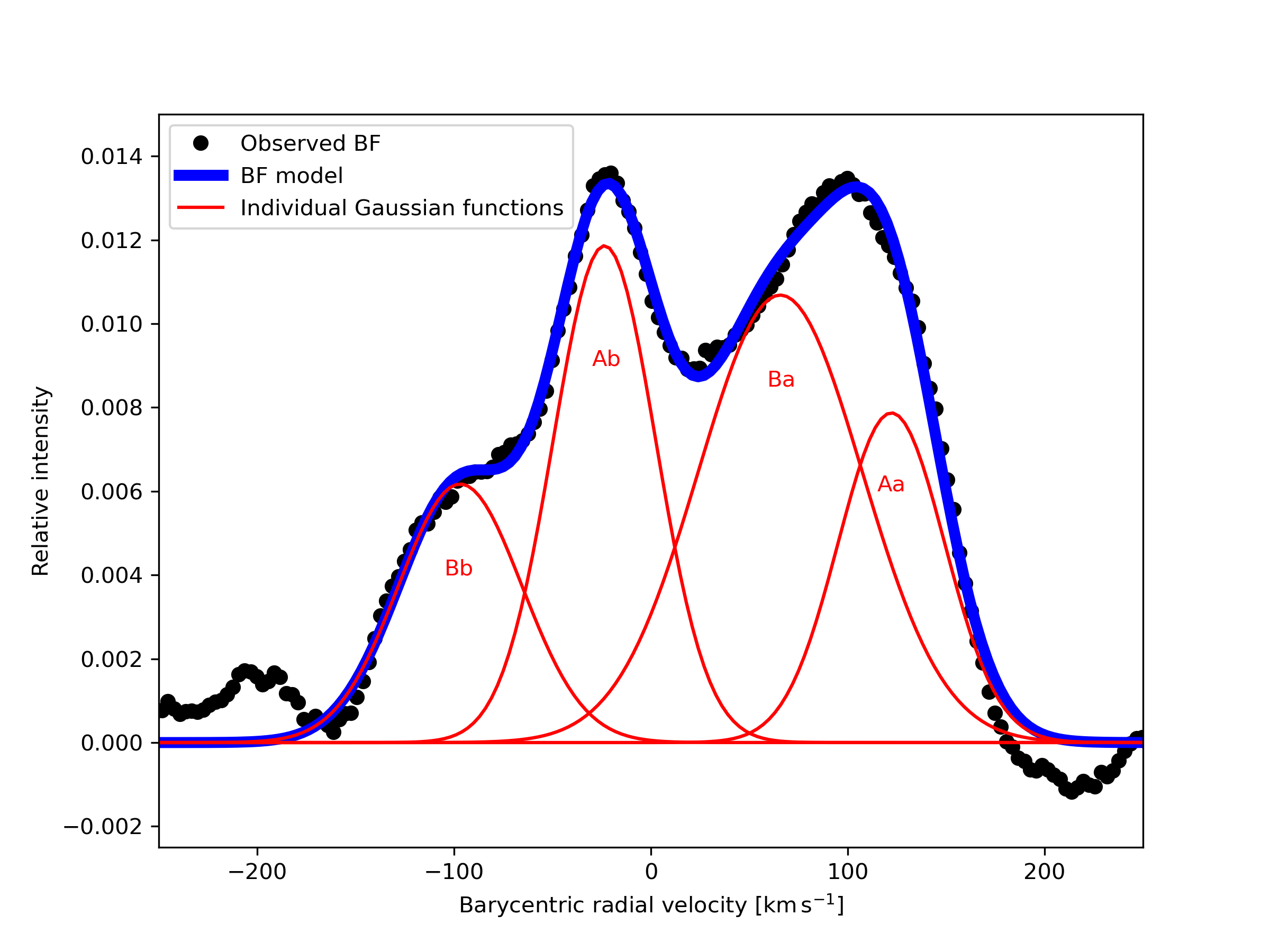}
   \caption{Example of a typical broadening function, in which the four components of BU CMi are relatively well-separated, and the corresponding model fit. Black points denote the BF calculated from the spectrum of BU CMi observed on March 26, 2021 at SP, the fitted Gaussian functions corresponding to the individual components of BU CMi are denoted with thin red lines while the wider blue line is the sum of these Gaussian functions.}
   \label{fig:BFplot} % Fig. 2
\end{figure}

We extracted the broadening functions from all spectra of BU CMi using the publicly available software BF-rvplotter\footnote{\href{https://github.com/mrawls/BF-rvplotter}{https://github.com/mrawls/BF-rvplotter}} in a $\pm$300\,km\,s$^{-1}$ velocity range using 3\,km\,s$^{-1}$ RV bins, and a Gaussian smoothing using a rolling window of five datapoints. We tried templates of different ({G-}, F-, A- spectral type) stars, and the spectrum of Vega with A0V spectral type produced the best quality BFs over the 4400-4600\,\AA\, and 4900-5300\,\AA\, wavelength range that we used for the calculations. The sum of four Gaussian functions was then fitted to the resulting broadening functions to find the peak positions of the profiles of the component stars in order to determine their RVs. Although the use of Gaussian functions is not the most appropriate choice for fitting the BFs of rapidly rotating stars, we found it to work well in the case of the BFs of BU CMi, as we were able to constrain satisfactory model fits to them using the sum of four Gaussian functions. An example of a typical BF along with the corresponding model fit is displayed in Fig.\,\ref{fig:BFplot}. The resulting RVs and their 1-sigma uncertainties are listed in Table~\ref{Tab:RVs}.

\begin{table*}
\centering
\caption{Radial velocities of BU~CMi determined by multi-component fitting to the broadening functions\label{tbl:rvs}.}
\scalebox{0.80}{\begin{tabular}{@{}lrrrrrrrrc}
\hline
BJD & RV$_\mathrm{Aa}$  & $\sigma_\mathrm{Aa}$ & RV$_\mathrm{Ab}$  & $\sigma_\mathrm{Ab}$  & RV$_\mathrm{Ba}$  & $\sigma_\mathrm{Ba}$ & RV$_\mathrm{Bb}$  & $\sigma_\mathrm{Bb}$ & Instr.\\          
$-2\,400\,000$ & km\,s$^{-1}$ & km\,s$^{-1}$ & km\,s$^{-1}$ & km\,s$^{-1}$ & km\,s$^{-1}$ & km\,s$^{-1}$ & km\,s$^{-1}$ & km\,s$^{-1}$ & \\
\hline
58926.33981 &-100.82 & 2.71  & 104.82 & 0.80  &  27.64 & 1.02 & -18.75 &  1.36 & SP\\
58927.34300 &  55.92 & 1.23  & -91.25 & 3.40  & -38.38 & 3.48 & 134.65 &  0.59 & SP\\
58928.34657 &  ...   & ...	 &  29.20 & 0.51  &  83.95 & 0.74 & -10.59 &  1.19 & SP\\
58944.26046 &-116.54 & 1.34  & 102.52 & 1.33  &  ...   & ...  &  62.74 & 8.23 & SP\\
58945.26827 & 106.37 & 4.67  &-161.82 & 2.98  & 168.35 & 2.91 & -68.04 &  1.89 & SP\\
58946.26321 & -74.88 & 4.40  &   9.27 & 5.00  & -42.30 & 4.28 & 155.31 &  0.75 & SP\\
58947.28005 &-125.27 & 4.28  & 111.43 &14.02  &   8.88 &13.36 & 145.91 &  2.74 & SP\\
58948.26070 & 120.95 & 2.74  &-153.75 & 1.27  & 177.05 & 2.77 & -31.64 &  0.79 & SP\\
58960.27911 &  44.48 & 1.93  &-129.69 & 1.21  & -19.71 & 1.69 & 133.07 &  1.41 & SP\\
58962.27609 & -69.44 & 3.07  &  39.89 & 2.20  &  77.87 & 2.82 &  12.76 & 12.20 & SP\\
59163.61701 &  ...   & ...	 &   0.16 &7.69  & 103.47 &5.28 & -78.91 &  3.65 & SP\\
59164.66476 & -97.24 & 2.07  & 172.32 & 1.17  & -33.89 & 0.80 &  62.87 &  1.02 & SP\\
59166.62999 &  ...   & ...	 &  ...   & ...   &  66.36 & 1.61 & -17.27 &  1.34 & SP\\
59177.65130 & 170.93 & 1.82  &-142.80 & 1.13  &  81.29 & 1.45 &  24.56 &  2.12 & SP\\
59178.61684 & -76.14 & 0.82  &  ...   & ...   & -11.33 & 1.40 & 124.14 & 6.93 & SP\\
59179.64615 &-113.25 & 3.27  &  89.70 & 2.14  & 145.01 & 1.90 & -30.68 &  1.37 & SP\\
59180.55889 & 116.90 & 3.17  &-155.45 & 2.19  & 171.34 & 3.33 & -53.21 &  1.49 & SP\\
59185.66472 & -69.51 & 3.60  &  71.14 & 9.80  &  32.97 &46.74 & 103.35 & 6.78 & SP\\
59196.56929 &-103.79 & 18.21 & 104.92 &6.95  & 198.88 & 3.73 & -53.46 &  2.69 & SP\\
59197.56304 & -20.41 & 7.37 &  37.01 &22.85  & -78.00 & 3.91 & 187.48 &  0.76 & SP\\
59203.53319 &  48.40 & 101.11&  31.69 & 0.71  & 143.25 &5.52 &  -8.35 &  2.00 & SP\\
59216.41157 &  59.33 & 18.02 & -22.99 & 4.46  & 175.59 &3.16  & -71.59 &5.46  & SP\\
59224.48614 & 129.40 & 0.65  &-135.62 & 3.17  & -27.01 &2.49  &  ...     & ... & SP\\
59226.43383 &-115.92 & 2.41  & 110.19 & 0.62  & ...	   &  ... & -23.44 & 1.29  & SP\\
59246.36404 &  ...   & ...	 & 163.93 & 1.44  & -81.19 & 3.10 &  49.22 & 0.87  & SP\\
59246.38986 &  ...   & ...	 & 170.57 & 1.50  & -86.03 & 3.68 &  54.69 & 0.81  & SP\\
59267.34294 & -96.01 & 1.89  & 173.43 & 0.76  &  20.93 & 3.76 & -25.99 & 2.88  & SP\\
59267.37076 & -93.53 & 1.80  & 171.72 & 0.78  &  -7.65 & 2.59 & -45.48 & 1.31  & SP\\
59268.35987 & 120.20 & 0.64  & -21.68 & 3.28  & ...	   & ...  &-147.59 & 3.52  & SP\\
59270.35148 & -72.73 & 2.48  & 135.05 & 1.60  & ...	   & ...  &   3.04 & 2.38  & SP\\
59271.36496 & 124.56 & 2.35  & -10.76 & 1.11  & ...	   & ...  & -87.70 & 0.77  & SP\\
59274.34879 & 111.83 & 1.00  & -77.05 & 1.22  & ...	   & ...  & -33.09 & 1.00  & SP\\
59275.39088 &  ...   & ...	 &   5.05 & 3.29  &  61.05 & 1.66 & -56.85 & 1.50  & SP\\
59282.38840 &-101.56 & 2.77  & 120.54 & 7.41 & -43.46 & 1.86 & 159.72 & 6.35 & SP\\
59284.28329 &  ...   & ...	 &  64.77 & 2.27  & ...	   &  ... & -88.06 & 0.72  & SP\\
59293.31352 & -28.53 & 17.69 & ...    &  ...  &  26.36 &5.30 &  77.70 & 4.64  & SP\\
59299.32690 & -91.48 & 1.34  &  58.53 & 0.94  & -39.68 & 3.19 & 133.56 & 0.51  & SP\\
59300.26297 & -96.87 & 2.29  &  65.73 &18.25  & 122.18 &5.46 & -23.58 & 1.76  & SP\\
59302.28482 &-111.92 & 1.65  &  77.20 & 1.37  & -55.80 & 2.82 & 194.55 & 0.98  & SP\\
59315.27384 &  -8.63 & 6.14 &  ...   & ...   & -75.11 &8.59 & 224.81 & 1.17  & SP\\
59563.49367 &  ...   & ...	 &  27.95 & 6.55 & -32.49 &6.99 & 180.45 & 5.68   & Konkoly\\
59564.58656 & -90.07 & 2.60	 &  ...   & ...   & 127.06 & 1.55 & -15.46 & 0.75    & Konkoly\\
59581.54165 & -46.64 & 1.50	 &  77.10 & 1.70  & 154.08 & 3.58 & -44.94 & 2.03    & Konkoly\\
59581.57345 & -56.10 & 1.73	 &  54.85 & 1.45  & 148.50 & 2.76 & -47.84 & 1.80    & Konkoly\\
59595.59930 &  56.80 & 1.16	 &  ...   & ...   & -58.90 & 3.50 & 139.73 & 3.81    & Rozhen\\
59596.53846 &  ...   & ...	 & 109.56 & 3.79  & -11.70 & 3.40 &  51.95 & 2.94    & Rozhen\\
59620.37490 &	1.40 & 18.21 & 185.91 & 4.54  &  70.04 &12.95 &-122.52 & 11.38   & SP\\
59621.42985 & 243.28 & 2.09	 & -88.02 & 0.89  & -31.71 & 1.32 &  16.14 & 2.13    & SP\\
59623.41741 & -77.78 & 1.14	 & 176.91 & 1.83  &  24.32 & 0.83 &-126.58 & 1.37    & SP\\
59624.46480 & 226.89 & 0.96	 & -84.41 & 1.95  &  49.31 & 1.27 & -54.77 & 2.04    & SP\\
59633.30758 & 193.35 & 35.03 &  ...   & ...   &  73.74 & 1.43 & -77.63 & 1.44    & Konkoly\\
59634.42208 & -46.82 & 11.64 & 151.91 & 3.83  &  -2.65 & 3.22 &  40.64 & 4.95    & Konkoly\\
59640.48490 & -47.96 & 12.68 &  94.67 & 2.32  & 143.33 & 4.52 & -95.93 & 3.75    & SP\\
59641.34036 & -57.00 & 2.25	 &  46.88 & 1.53  &-140.68 & 1.57 & 136.00 & 0.80    & SP\\
59649.42080 & -56.00 & 2.01	 & 125.09 & 2.78  &  75.80 & 1.57 & -15.53 & 1.22    & SP\\
59650.32442 & -34.60 & 1.13	 &  31.63 & 1.17  &  64.99 & 4.68 & -83.83 & 0.94    & SP\\
59651.32302 &  67.72 & 1.51	 & -76.36 & 1.08  &-136.84 & 2.18 & 166.05 & 1.17    & SP\\
59652.34238 & -93.63 & 3.54	 &  60.93 & 5.94 &  38.51 & 4.65 &   0.22 & 3.09    & SP\\
59653.36525 &  34.33 & 1.25	 & -28.27 & 3.26  &  80.51 & 1.64 & -84.74 & 2.08    & SP\\
\hline
\end{tabular}}
 \label{Tab:RVs}
\end{table*}

\addtocounter{table}{-1}
\begin{table*}
\centering
\caption{\textit{continued}}
\scalebox{0.80}{\begin{tabular}{@{}lrrrrrrrrc}
\hline
BJD & RV$_\mathrm{Aa}$  & $\sigma_\mathrm{Aa}$ & RV$_\mathrm{Ab}$  & $\sigma_\mathrm{Ab}$  & RV$_\mathrm{Ba}$  & $\sigma_\mathrm{Ba}$ & RV$_\mathrm{Bb}$  & $\sigma_\mathrm{Bb}$ & Instr.\\          
$-2\,400\,000$ & km\,s$^{-1}$ & km\,s$^{-1}$ & km\,s$^{-1}$ & km\,s$^{-1}$ & km\,s$^{-1}$ & km\,s$^{-1}$ & km\,s$^{-1}$ & km\,s$^{-1}$ & \\
\hline
59657.27741  &  69.05	& 0.66   & -54.75   & 2.12  & ...      & ...	& ...	   & ...   &  SP\\  
59658.36099  &  ...	& ...	 &  ...     & ...   & -23.32   & 0.67	& 126.93   & 2.00  &  SP\\
59658.42893  & -28.92	& 31.67  & 126.78   & 3.64  & -20.98   & 5.00	& ...	   & ...   &  Konkoly\\
59659.42000  & 100.44	& 7.05   &-110.08   & 1.74  & 167.04   & 4.86	& -39.10   & 2.44  &  Konkoly\\
59660.44895  &  ...	& ...	 &  ...     & ...   &  62.28   & 1.41	&  27.40   & 4.37  &  Konkoly\\
59661.42361  & -82.93	& 4.98   &  94.62   & 4.45  & -19.90   &5.80	& 164.64   & 2.63  &  Konkoly\\
59662.41581  & 143.85	& 2.84   &-114.05   & 6.90  &  94.24   & 1.22	&   7.58   & 5.26  &  Konkoly\\
59666.36259  &  -4.57	& 1.01   &  -4.57   & 1.01  & 174.67   &6.22	& ...	   & ...   &  SP\\
59682.29860  & -91.78	& 8.55   &  -6.41   & 8.27  & 155.33   & 1.13	& -29.63   & 1.71  &  SP\\
59699.32743  & -61.54	& 2.29   &  96.99   & 1.77  & 142.89   & 1.96	& -30.28   & 1.84  &  SP\\
59899.85591  & -82.11	& 12.45  &  82.76   & 20.10 & -18.40   & 6.16	& 133.04   & 3.44  &  CTIO\\
59903.82298  & 153.29	& 17.20  &-121.13   & 5.33  & 104.08   & 6.94	&   5.75   & 3.70  &  CTIO\\
59906.83041  & 116.27	& 7.29   &-109.47   & 2.87  & ...      & ...	&  37.10   & 2.93  &  CTIO\\
59916.79609  & -89.13	& 6.55   &  31.93   & 9.85  & 128.10   & 9.98	& -26.84   & 8.04  &  CTIO\\
59919.80506  & -105.34  & 1.95   &  ...     & ...   &  72.90   &  4.96  &    32.76  &  4.36 &  CTIO\\
59922.78577  & -101.93  & 4.21   &   74.87  & 2.33  &  ...     & ...    &    ...    &  ...  &  CTIO\\
59924.78928  & ...      & ...    &  -94.14  & 5.02  &  61.96   &  3.02  &    ...    &  ...  &  CTIO\\
59936.74125  & ...      & ...    &  -9.78   & 2.54  & 133.53   &  6.08  &   -52.85  &  1.19 &  CTIO\\
59949.76298  & ...      & ...    &  ...     & ...   & 122.66   &  1.70  &   -52.32  &  1.40 &  CTIO\\
59952.71053  & ...      & ...    &  ...     & ...   &  73.27   &  3.00  &   -29.47  &  0.95 &  CTIO\\
59955.69837  &  -73.16  & 3.19   &   81.66  & 7.63  &  48.93   &  2.36  &     3.66  &  1.58 &  CTIO\\
59998.58934  & ...      & ...    &   46.37  & 4.29  &  100.26  &  4.31  &   -59.53  &  7.28 &  CTIO\\
60000.60639  & ...      & ...    & -139.58  & 3.59  &  -86.07  &  4.60  &   156.41  &  1.03 &  CTIO\\
60013.61812  & ...      & ...    &   85.54  & 4.79  &  -73.46  &  0.74  &   154.44  &  1.81 &  CTIO\\
60013.63551  & ...      & ...    &   81.75  & 5.98  &  -72.99  &  0.86  &   152.34  &  2.08 &  CTIO\\
60019.56255  &  -44.48  & 1.29   &  100.68  & 2.47  &  -13.41  &  3.62  &   141.85  &  1.79 &  CTIO\\
60033.48258  &   32.96  & 1.34   &  ...     & ...   &  -45.16  &  1.37  &   122.10  &  0.67 &  CTIO\\
60065.30014  & ...      & ...    &  -146.19 & 6.79  &  -13.56  &  3.24  &   146.15  &  2.90 &  CTIO\\
60065.31060  & ...      & ...    &  -157.82 & 3.98  &  -22.58  &  3.01  &   167.87  &  6.52 &  CTIO\\
60065.32105  & ...      & ...    &  -148.62 & 7.84  &  -16.11  &  3.51  &   154.59  &  6.43 &  Konkoly\\
\hline
\end{tabular}}

{\em Note.} RV points newer than BJD 2\,459\,917 were not used for the spectro-photodynamical analysis.
\end{table*}

\section{Period study}
\label{sec:Period_study}

\begin{figure*}
    \centering
    \includegraphics[width=0.48\linewidth]{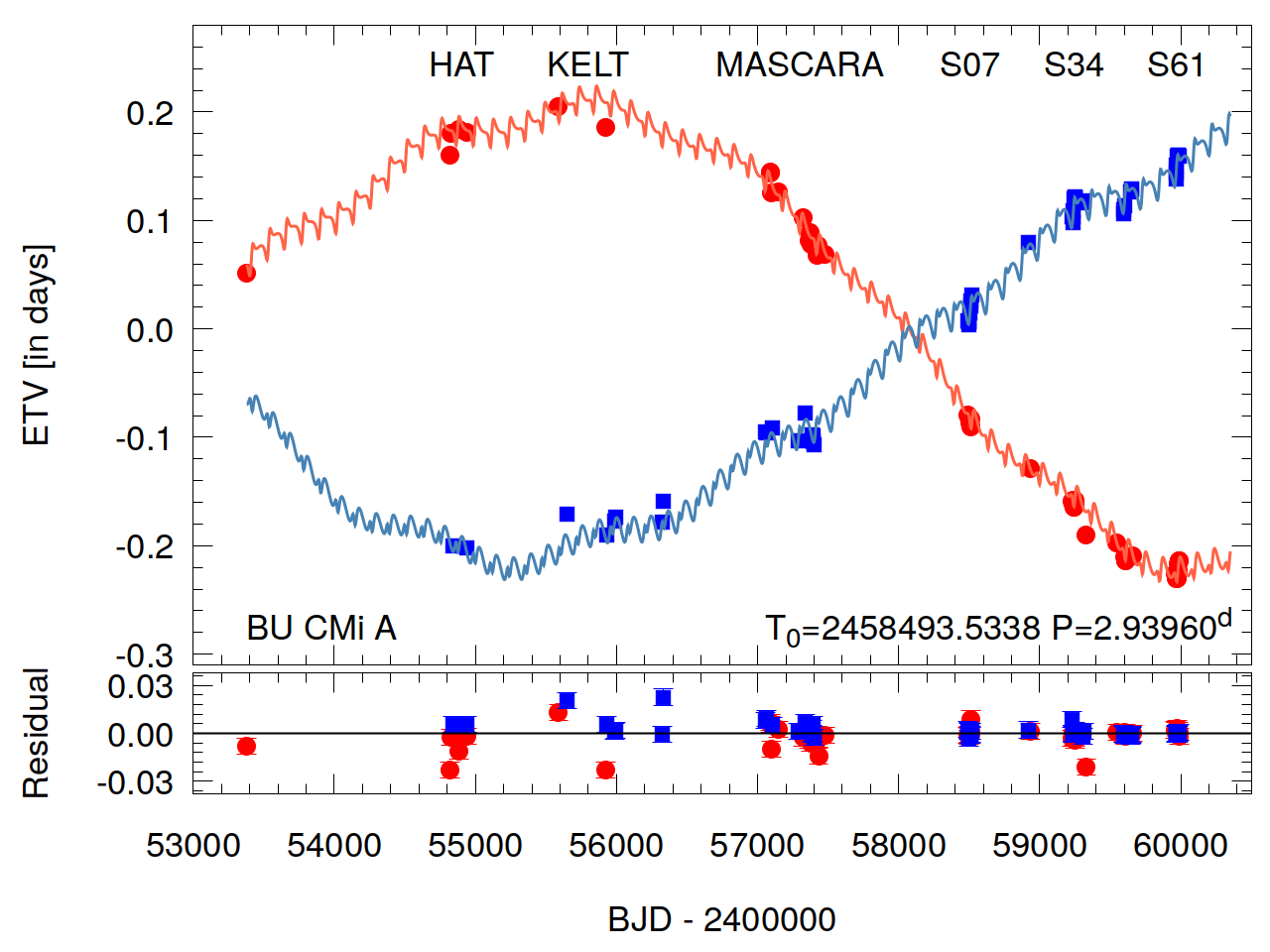}\includegraphics[width=0.48\linewidth]{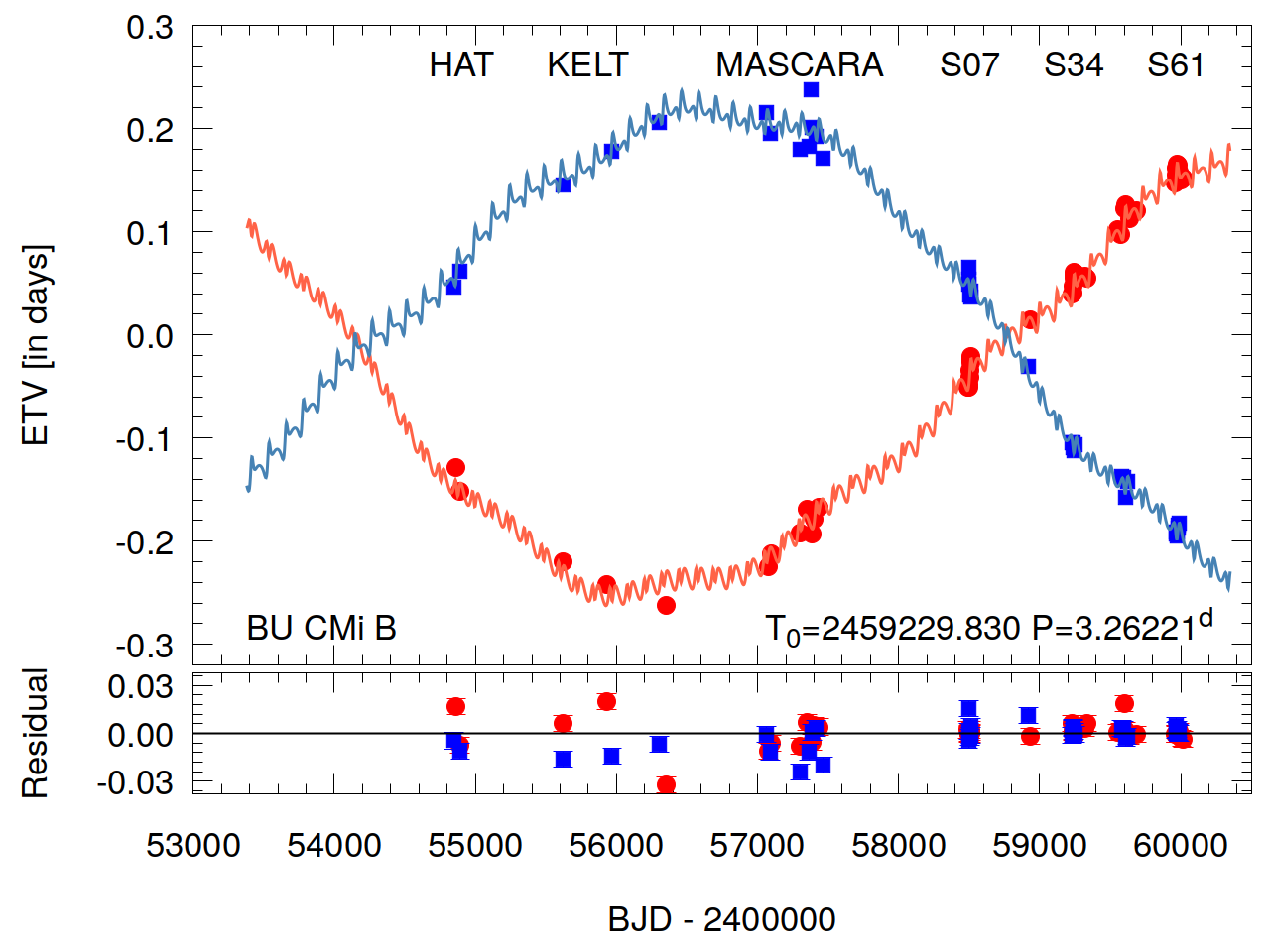}
    \includegraphics[width=0.48\linewidth]{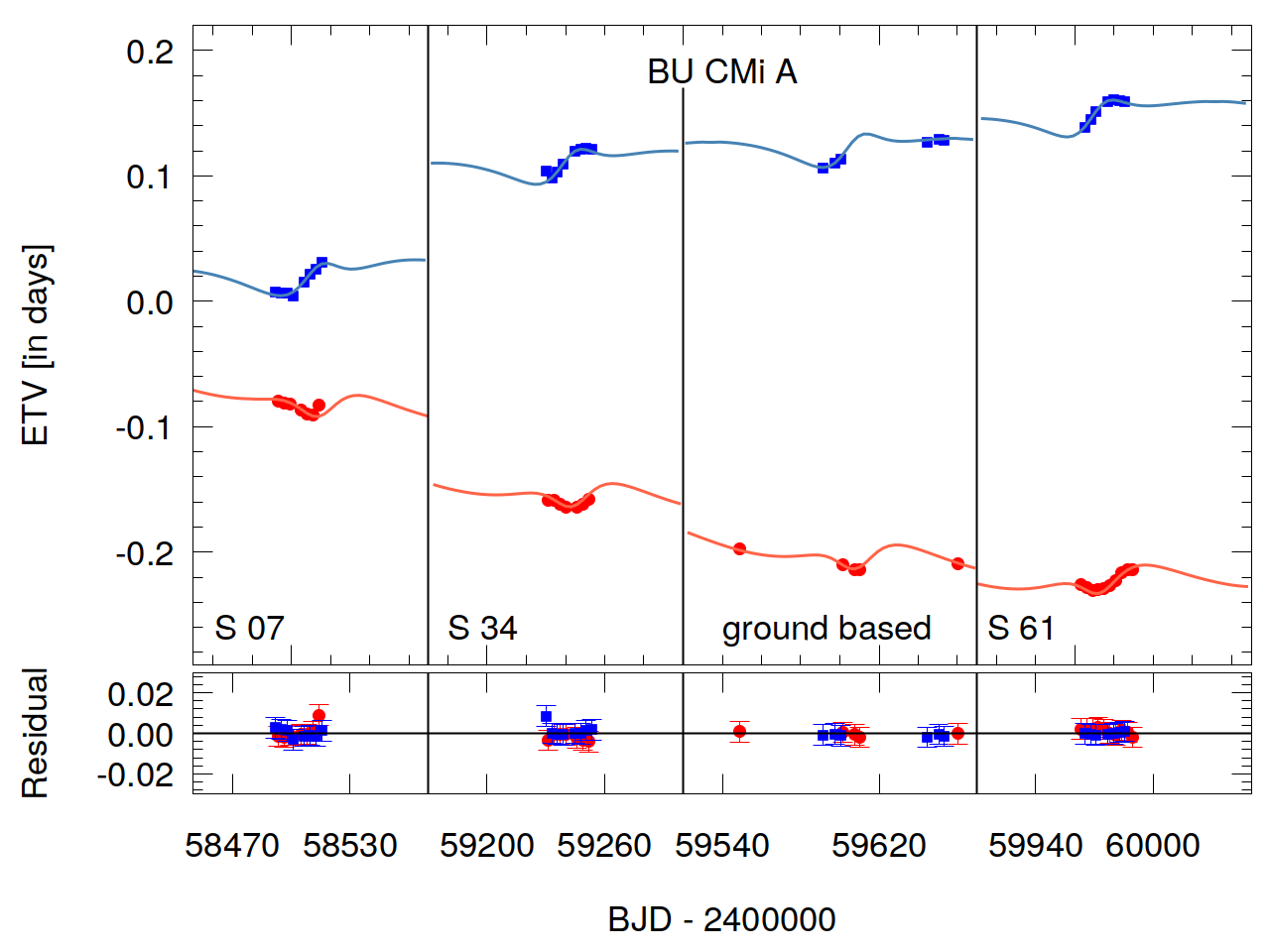}\includegraphics[width=0.48\linewidth]{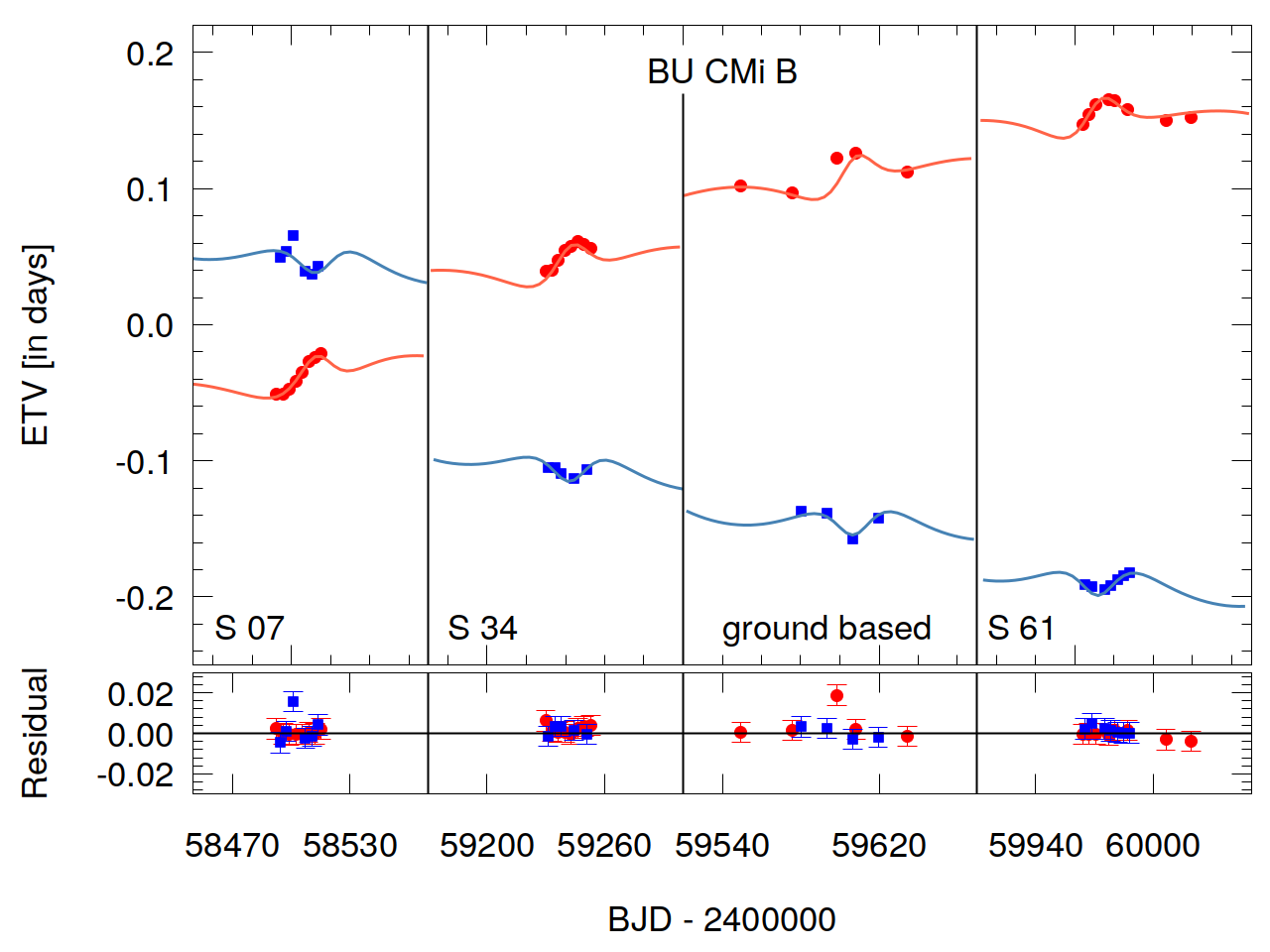}
    \caption{The primary (red) and secondary (blue) ETV curves for binaries A and B (left and right, respectively). The larger symbols represent the measured values, while the smaller points connected with straight lines stand for the photodynamical model values, obtained from numerical integration of the orbital motions. The bottom panels are zoom-ins around the three \textit{TESS} sectors and one set of ground-based observations.}
   \label{fig:etv}
\end{figure*} % Fig. 3

\subsection{Determination of times of minima}
\label{sec:ToMs}

In order to carry out a preliminary period study with the usual method of analyzing the eclipse timing variations (ETV), we determined accurate mid-eclipse times from the various photometric light curves. Due to the presence of frequently and strongly overlapping eclipses in the two EBs, this process required some extra care.

In the case of the quasi-continuous, 25--26-day-long \textit{TESS} observations, we disentangled the light curves of the two EBs, and mutually removed the signals of the other EB from the light curve in the same manner as was also done for our analyses of V994~Her, \citep{zascheetal23}.  This needs to be done in order to avoid losing the overlapping eclipse events and, moreover, to enhance the accuracies of those times where the eclipses of the two EBs do not overlap each other, but other light-curve features (e.g. the periastron bumps) of one of the binaries distorts the eclipsing signal of the other EB.  Due to the rapid and large amplitude ETVs, this method was done separately for the three sectors and, even then, in the case of a few fully overlapping eclipsing events, the disentanglement was not perfect. As a result, we had to drop out 2-3 outlier eclipse times from our analysis.

In the absence of well-covered out-of-eclipse light curves for our 2021--2023 targeted ground-based follow-up eclipse observations, we were unable to carry out the same disentanglement process, and had no choice but to simply exclude a few blended eclipses from our ETV analysis.

In the case of the \textit{MASCARA}, \textit{KELT} and \textit{HAT} data, after their conversion from the original Heliocentric Julian Date (HJD) times to Barycentric Julian Date (BJD), we again disentangled the two binaries' signals from the light curves.  We then took those eclipse observations where both the ingress and egress portions of the same eclipses were measured, and we determined mid-eclipse times for these individual eclipses. The resulting mid-eclipse times as determined from the \textit{TESS}, \textit{KELT}, and \textit{MASCARA} archival data, as well as our new ground-based observations, are listed in Tables~\ref{Tab:BU_CMi_A_(TIC_271204362A)_ToM}, \ref{Tab:BU_CMi_B_(TIC_271204362B)_ToM}.

Finally, we utilized some additional published eclipse times from the papers of \citet{ibvs} and \citet{2021ARep...65..826V}.  Regarding this latter work, \citet{2021ARep...65..826V} tabulate their own eclipse times determined from the \textit{MASCARA} data and the \textit{TESS} sector 7 and 34 observations. We do not use those particular times, as we have determined our own eclipse times from the same datasets using our own method.  We do use, however, most of the eclipse times that \citet{2021ARep...65..826V} determined from their own observations, and to which we do not have access.  Unfortunately, they do not provide uncertainties for these times, so we arbitrarily take an estimated error of 0.0001\,days for each of these data points.

The combination of all these data has provided 102 and 81 mid-eclipse times for binaries A and B, respectively.

\begin{table*}
\centering
\caption{Eclipse times for BU CMi A}
 \label{Tab:BU_CMi_A_(TIC_271204362A)_ToM}
\begin{tabular}{@{}lrlllrlllrll}
\hline
BJD & Cycle  & std. dev. & ref & BJD & Cycle  & std. dev. & ref & BJD & Cycle  & std. dev. & ref\\ 
$-2\,400\,000$ & no. &   \multicolumn{1}{c}{$(d)$} & & $-2\,400\,000$ & no. &   \multicolumn{1}{c}{$(d)$} & & $-2\,400\,000$ & no. &   \multicolumn{1}{c}{$(d)$} & \\ 
\hline
53378.68123 & -1740.0 & 0.00130 & 1 & 57398.42545 &  -372.5 & 0.00111 & 5 &  59250.60298 &   257.5 & 0.00007  &  6 \\ 
54822.13354 & -1249.0 & 0.00154 & 2 & 57423.58727 &  -364.0 & 0.00158 & 5 &  59251.79253 &   258.0 & 0.00005  &  6 \\ 
54825.09356 & -1248.0 & 0.00031 & 2 & 57429.47566 &  -362.0 & 0.00025 & 5 &  59253.54149 &   258.5 & 0.00007  &  6 \\ 
54840.88100 & -1242.5 & 0.00066 & 2 & 57432.40776 &  -361.0 & 0.00172 & 5 &  59309.38877 &   277.5 & 0.00011  &  7 \\ 
54880.94923 & -1229.0 & 0.00069 & 2 & 57479.44068 &  -345.0 & 0.00037 & 5 &  59312.33064 &   278.5 & 0.00029  &  7 \\ 
54936.79939 & -1210.0 & 0.00060 & 2 & 58492.07146 &    -0.5 & 0.00056 & 6 &  59331.12995 &   285.0 & 0.00010  &  4 \\ 
54940.82537 & -1208.5 & 0.00050 & 2 & 58493.45400 &     0.0 & 0.00023 & 6 &  59548.65262 &   359.0 & 0.00009  &  7 \\ 
55589.41391 &  -988.0 & 0.00120 & 3 & 58495.00998 &     0.5 & 0.00108 & 6 &  59591.58024 &   373.5 & 0.00012  &  7 \\ 
55649.29964 &  -967.5 & 0.00766 & 3 & 58496.39232 &     1.0 & 0.00036 & 6 &  59597.46355 &   375.5 & 0.00011  &  7 \\ 
55924.50904 &  -874.0 & 0.13849 & 3 & 58497.94961 &     1.5 & 0.00405 & 6 &  59600.40685 &   376.5 & 0.00011  &  7 \\ 
55928.54252 &  -872.5 & 0.00054 & 3 & 58499.33127 &     2.0 & 0.00016 & 6 &  59601.55295 &   377.0 & 0.00010  &  7 \\ 
55987.34743 &  -852.5 & 0.00137 & 3 & 58500.88700 &     2.5 & 0.00047 & 6 &  59607.42797 &   379.0 & 0.00013  &  7 \\ 
55993.23042 &  -850.5 & 0.00010 & 4 & 58505.20550 &     4.0 & 0.00034 & 6 &  59610.36790 &   380.0 & 0.00020  &  7 \\ 
56325.40060 &  -737.5 & 0.00102 & 3 & 58506.77692 &     4.5 & 0.00056 & 6 &  59644.51396 &   391.5 & 0.00020  &  7 \\ 
56331.29927 &  -735.5 & 0.00649 & 3 & 58508.14199 &     5.0 & 0.00015 & 6 &  59650.39532 &   393.5 & 0.00011  &  7 \\ 
57057.44400 &  -488.5 & 0.00032 & 5 & 58509.72284 &     5.5 & 0.00037 & 6 &  59653.33407 &   394.5 & 0.00009  &  7 \\ 
57060.38319 &  -487.5 & 0.00078 & 5 & 58511.08072 &     6.0 & 0.00016 & 6 &  59660.34597 &   397.0 & 0.00013  &  7 \\ 
57091.48893 &  -477.0 & 0.00036 & 5 & 58512.66692 &     6.5 & 0.00048 & 6 &  59963.10789 &   500.0 & 0.00001  &  6 \\ 
57094.42835 &  -476.0 & 0.00029 & 5 & 58514.02806 &     7.0 & 0.00077 & 6 &  59964.94229 &   500.5 & 0.00001  &  6 \\ 
57097.34970 &  -475.0 & 0.00241 & 5 & 58515.61193 &     7.5 & 0.00064 & 6 &  59966.04491 &   501.0 & 0.00001  &  6 \\ 
57107.42078 &  -471.5 & 0.00074 & 5 & 58921.32496 &   145.5 & 0.00010 & 4 &  59967.88775 &   501.5 & 0.00001  &  6 \\ 
57144.38355 &  -459.0 & 0.00124 & 5 & 58934.34520 &   150.0 & 0.00010 & 4 &  59968.98246 &   502.0 & 0.00001  &  6 \\ 
57286.72438 &  -410.5 & 0.00030 & 5 & 59230.00713 &   250.5 & 0.00009 & 6 &  59970.83371 &   502.5 & 0.00001  &  6 \\ 
57323.67544 &  -398.0 & 0.00045 & 5 & 59231.21430 &   251.0 & 0.00006 & 6 &  59971.92238 &   503.0 & 0.00001  &  6 \\ 
57336.72404 &  -393.5 & 0.00501 & 5 & 59232.94169 &   251.5 & 0.00007 & 6 &  59974.86321 &   504.0 & 0.00001  &  6 \\ 
57367.74872 &  -383.0 & 0.00057 & 5 & 59234.15447 &   252.0 & 0.00006 & 6 &  59976.72136 &   504.5 & 0.00000  &  6 \\ 
57370.69204 &  -382.0 & 0.00039 & 5 & 59235.88604 &   252.5 & 0.00006 & 6 &  59977.80493 &   505.0 & 0.00001  &  6 \\ 
57373.63485 &  -381.0 & 0.00141 & 5 & 59237.09115 &   253.0 & 0.00005 & 6 &  59979.66193 &   505.5 & 0.00000  &  6 \\ 
57376.56410 &  -380.0 & 0.00025 & 5 & 59238.83148 &   253.5 & 0.00007 & 6 &  59980.74840 &   506.0 & 0.00001  &  6 \\ 
57379.50832 &  -379.0 & 0.00083 & 5 & 59240.02798 &   254.0 & 0.00005 & 6 &  59982.60139 &   506.5 & 0.00001  &  6 \\ 
57386.67631 &  -376.5 & 0.00025 & 5 & 59244.72129 &   255.5 & 0.00007 & 6 &  59983.69460 &   507.0 & 0.00001  &  6 \\ 
57389.61151 &  -375.5 & 0.00020 & 5 & 59245.90716 &   256.0 & 0.00005 & 6 &  59985.54002 &   507.5 & 0.00001  &  6 \\ 
57392.54958 &  -374.5 & 0.00047 & 5 & 59247.66224 &   256.5 & 0.00008 & 6 &  59986.63671 &   508.0 & 0.00005  &  6 \\ 
57395.49440 &  -373.5 & 0.00021 & 5 & 59248.84941 &   257.0 & 0.00005 & 6 &  59989.57610 &   509.0 & 0.00015  &  6 \\ 
\hline
\end{tabular}

{\em Notes.} Integer and half-integer cycle numbers refer to primary and secondary eclipses, respectively. \\
References: 1: \citet{ibvs}; 2: \textit{HAT} - this paper; 3: \textit{KELT} - this paper; 4: \citet{2021ARep...65..826V}; 5: \textit{MASCARA} - this paper; 6: \textit{TESS} - this paper; 7: Ground-based follow up - this paper.
\end{table*}

\begin{table*}
\centering
\caption{Eclipse times for BU CMi B}
\begin{tabular}{@{}lrlllrlllrll}
\hline
BJD & Cycle  & std. dev. & ref & BJD & Cycle  & std. dev. & ref & BJD & Cycle  & std. dev. & ref\\ 
$-2\,400\,000$ & no. &   \multicolumn{1}{c}{$(d)$} & & $-2\,400\,000$ & no. &   \multicolumn{1}{c}{$(d)$} & & $-2\,400\,000$ & no. &   \multicolumn{1}{c}{$(d)$} & \\ 
\hline
54847.09728 & -1343.5 & 0.00043 & 2 & 58495.78182 &  -225.0 & 0.00065 & 6 & 59321.22889 &    28.0 & 0.00010  & 4 \\ 
54864.86467 & -1338.0 & 0.00087 & 2 & 58497.51748 &  -224.5 & 0.00097 & 6 & 59334.27590 &    32.0 & 0.00010  & 4 \\ 
54890.93880 & -1330.0 & 0.00202 & 2 & 58499.04739 &  -224.0 & 0.00064 & 6 & 59549.62826 &    98.0 & 0.00013  & 7 \\ 
54892.78392 & -1329.5 & 0.00062 & 2 & 58500.79134 &  -223.5 & 0.00167 & 6 & 59575.72105 &   106.0 & 0.00017  & 7 \\ 
55618.34394 & -1107.0 & 0.00130 & 3 & 58502.31563 &  -223.0 & 0.00054 & 6 & 59580.38070 &   107.5 & 0.00027  & 7 \\ 
55620.34039 & -1106.5 & 0.00078 & 3 & 58505.58460 &  -222.0 & 0.00056 & 6 & 59593.42771 &   111.5 & 0.00015  & 7 \\ 
55931.49403 & -1011.0 & 0.00370 & 3 & 58507.28947 &  -221.5 & 0.00064 & 6 & 59598.58237 &   113.0 & 0.00009  & 7 \\ 
55969.42899 &  -999.5 & 0.00308 & 3 & 58508.85441 &  -221.0 & 0.00034 & 6 & 59606.45774 &   115.5 & 0.00025  & 7 \\ 
56305.46439 &  -896.5 & 0.00070 & 3 & 58510.54976 &  -220.5 & 0.00037 & 6 & 59608.37210 &   116.0 & 0.00018  & 7 \\ 
56352.29850 &  -882.0 & 0.00329 & 3 & 58512.12004 &  -220.0 & 0.00078 & 6 & 59619.52224 &   119.5 & 0.00017  & 7 \\ 
57065.56905 &  -663.5 & 0.00020 & 5 & 58513.81759 &  -219.5 & 0.00041 & 6 & 59634.45631 &   124.0 & 0.00011  & 7 \\ 
57076.54707 &  -660.0 & 0.00059 & 5 & 58515.38485 &  -219.0 & 0.00043 & 6 & 59683.39750 &   139.0 & 0.00033  & 7 \\ 
57088.38436 &  -656.5 & 0.00148 & 5 & 58921.52094 &   -94.5 & 0.00010 & 4 & 59963.97419 &   225.0 & 0.00001  & 6 \\ 
57099.39424 &  -653.0 & 0.00087 & 5 & 58936.24555 &   -90.0 & 0.00010 & 4 & 59965.26725 &   225.5 & 0.00001  & 6 \\ 
57301.67223 &  -591.0 & 0.00366 & 5 & 59229.86946 &     0.0 & 0.00008 & 6 & 59967.24418 &   226.0 & 0.00001  & 6 \\ 
57303.67503 &  -590.5 & 0.04406 & 5 & 59231.35608 &     0.5 & 0.00005 & 6 & 59968.52795 &   226.5 & 0.00001  & 6 \\ 
57350.62777 &  -576.0 & 0.00027 & 5 & 59233.13240 &     1.0 & 0.00007 & 6 & 59970.51363 &   227.0 & 0.00001  & 6 \\ 
57365.65974 &  -571.5 & 0.00160 & 5 & 59234.61858 &     1.5 & 0.00005 & 6 & 59975.05015 &   228.5 & 0.00002  & 6 \\ 
57378.76387 &  -567.5 & 0.00037 & 5 & 59236.40158 &     2.0 & 0.00008 & 6 & 59977.04164 &   229.0 & 0.00001  & 6 \\ 
57386.48796 &  -565.0 & 0.00036 & 5 & 59237.87619 &     2.5 & 0.00006 & 6 & 59978.31580 &   229.5 & 0.00001  & 6 \\ 
57388.51306 &  -564.5 & 0.00092 & 5 & 59239.67121 &     3.0 & 0.00009 & 6 & 59980.30316 &   230.0 & 0.00001  & 6 \\ 
57399.55166 &  -561.0 & 0.00035 & 5 & 59242.93668 &     4.0 & 0.00008 & 6 & 59981.58193 &   230.5 & 0.00002  & 6 \\ 
57414.60233 &  -556.5 & 0.00037 & 5 & 59244.39713 &     4.5 & 0.00006 & 6 & 59984.84735 &   231.5 & 0.00001  & 6 \\ 
57435.44751 &  -550.0 & 0.00029 & 5 & 59246.20221 &     5.0 & 0.00009 & 6 & 59986.82065 &   232.0 & 0.00002  & 6 \\ 
57463.51449 &  -541.5 & 0.00235 & 5 & 59249.46257 &     6.0 & 0.00007 & 6 & 59988.11137 &   232.5 & 0.00001  & 6 \\ 
58492.51983 &  -226.0 & 0.00044 & 6 & 59250.92821 &     6.5 & 0.00006 & 6 & 60006.38631 &   238.0 & 0.00009  & 7 \\ 
58494.25107 &  -225.5 & 0.00107 & 6 & 59252.72148 &     7.0 & 0.00008 & 6 & 60019.43692 &   242.0 & 0.00012  & 7 \\
\hline
\end{tabular}
 \label{Tab:BU_CMi_B_(TIC_271204362B)_ToM}

 {\em Notes.} Integer and half-integer cycle numbers refer to primary and secondary eclipses, respectively. \\
References: 1: \citet{ibvs}; 2: \textit{HAT} - this paper; 3: \textit{KELT} - this paper; 4: \citet{2021ARep...65..826V}; 5: \textit{MASCARA} - this paper; 6: \textit{TESS} - this paper; 7: Ground-based follow up - this paper.
\end{table*}

\subsection{Preliminary ETV study}
\label{sec:ETV_study}

The ETV curves formed from all the data listed in Tables~\ref{Tab:BU_CMi_A_(TIC_271204362A)_ToM} and \ref{Tab:BU_CMi_B_(TIC_271204362B)_ToM} are shown in Fig.~\ref{fig:etv}.  The top two panels show the overall ETV curves spanning all the data.  The left and right panels are for binary A and binary B, respectively.  The bottom panels are zoom-ins around the three \textit{TESS} sectors and one ground-based segment.

A first visual inspection of the ETV curves of both binaries shows two very remarkable features. First, each EB exhibits a large sinusoid with an amplitude of $\sim\pm 0.2$ days, and periods of $25-30$ years.  Note that the corresponding curves for the primary and secondary eclipses are anticorrelated.  This clearly indicates apsidal motion of the eccentric EBs.  We will show in Sect.~\ref{sect:photodynamics} that this apsidal motion is driven by the presence of the `other' binary.  

Secondly, the high-quality modern ETV points determined from the \textit{TESS} and ground-based follow-up observations exhibit clear $\sim120$-day periodic variations, with amplitudes of $30-40$ minutes. The shapes of these short-period ($\sim$120-day) ETVs clearly resemble the dynamically driven ETVs of several other tight and compact triple and quadruple star systems \citep[see, e.g.][]{borkovitsetal15,borkovitsetal16,borkovitsetal20,kostovetal21,kostovetal23}. We thereby associate the $\sim$120-day ETV features with the outer orbital period of the quadruple in which binary A orbits binary B.  Moreover, the apsidal motion periods are in perfect agreement with periods which are theoretically expected for tight triple systems with the observed inner and outer periods. It was therefore clear to us that BU CMi is an exceptionally compact doubly-eclipsing 2+2 hierarchical quadruple system.
%\footnote{When we identified this system as a potentially very interesting quadruple in 2021, we were unaware of the work presented in \citet{2021ARep...65..826V}.}

As a first attempt to understand these ETVs quantitatively, we made analytic ETV fits using the code and method described in \citet{borkovitsetal15}. As this code was developed for tight triple systems, unfortunately, it is unable to simultaneously fit the ETVs of both binaries. On the other hand, however, the analytical descriptions of the $P_\mathrm{out}$-timescale, as well as the apse-node perturbation terms, are included in the software package and, hence, it is very useful for getting reliable third-body solutions for dynamically-dominated ETV curves. When used to analyze the BU CMi ETV curves, the solutions confirmed the $P_\mathrm{out}\approx121.5$\,d outer orbital period and, moreover, the fact that the observed rapid apsidal motions are driven by the third-body perturbations of the other EB.\footnote{The period of the apsidal motion is not merely a free, adjustable parameter, but is calculated from the fitted inner and outer periods, mass ratios, and eccentricities as per theoretical formulae \citep[see][Appendix C for details]{borkovitsetal15}.} While our analytic ETV fits confirmed our initial hypothesis about the system configuration, the obtained fit was surprisingly poor for both binaries. This inadequate fit is due to an `extra feature' in the ETV curve with a period of $\sim$1000-1200 days that can be seen by casual inspection of the curves.

We were able to improve our analytical ETV fit substantially, however, when we ``added'' an additional, more distant fourth (i.e., in the present situation, fifth) stellar component to the system. A recent improvement of our analytic ETV software package makes it possible to fit simultaneously a second light-travel time effect (LTTE). We found that adding such a second LTTE component with a period of $P_\mathrm{outermost}\approx1100$\,days results in statistically excellent fits for the ETV curves of both binaries.  We emphasize, however, that this was a solution that was useful in the context of the fitting code, rather than having any physical significance. Such an extremely tight 2+2+1 quintuple system would be at best marginally stable; furthermore, the strong gravitational perturbations of the outermost component should also have to be taken into account. Later, we demonstrate, when discussing the fully consistent, photodynamical treatment (see Sect.~\ref{sect:photodynamics}), that the explanation for the origin of the extra cyclic terms in the ETVs with a $\approx1000-1200$-day period does not require the presence of an additional, more distant stellar component in the system. Rather, we show that the extra cyclic feature in the ETV curve arises naturally from the mutual gravitational perturbations between the two binaries.

Finally, we note that the results of this preliminary, analytic ETV analysis were used only for finding reliable input parameters for the complex, photodynamical fitting procedure (Sect.~\ref{sect:photodynamics}). Thus, a more detailed, quantitative discussion of the ETV results will be given later, in Sect.~\ref{sec:discussion}, in the context of the photodynamical results.

\section{Direct fitting to the spectra and \textit{TESS} light curves}
\label{sec:Theo_analyses}

The ETV analysis presented in section~\ref{sec:ETV_study} strongly indicates that the outer orbit of the system is very tight, with $P_\mathrm{out}\approx121.5$\,d. This means that the mutual orbit of the eclipsing binaries is relatively fast, resulting in EB center of mass radial velocity semi-amplitudes on the order of tens of km\,s$^{-1}$. Such a large RV variability superposed on EB orbits of a few days should be easy to detect.

The analysis of the spectra of BU~CMi is, however, significantly complicated by the fact that all its stellar components possess an early spectral type.  The spectra are therefore dominated by strong, wide hydrogen Balmer lines; these lines are very difficult to work with, as we cannot reliably detect their splitting and robustly calculate radial velocities of the individual stars in the system. On the other hand, the metal lines of the system are rather shallow; as a result, a relatively high signal-to-noise ratio is necessary to use these in our analysis. The line splitting due to the orbital motion in both pairs is clearly visible in the strongest metal line (Mg {\sc ii} 4481 \AA). Because the rotational velocities of the components ($v\,\sin\,i\sim$\,50--80 km\,s$^{-1}$) are approximately half the radial velocity semi-amplitudes ($K \sim $ 120--130 km\,s$^{-1}$), the components' profiles are almost always somewhat blended. Thus, a direct measurement of the radial velocity is rather difficult and may yield inaccurate results (see Section ~\ref{sec:RVs}).

Consequently, we attempted to directly model the observed spectra, under the assumption of two binary stars in a Keplerian orbit about each other. Two wavelength ranges were found to be appropriate: a blue region around the Mg {\sc ii} 4481 \AA~ line, covering 4380 - 4605.4 \AA, and a green region around the Mg {\sc i} triplet, covering 4900 - 5379.6 \AA. The spectra were weighted using the signal-to-noise ratios provided by the reduction pipeline at both 4400 \AA~ and 5500 \AA. The standard deviation at either of the spectral ranges was assumed to be 1/SNR\footnote{Only SP and CTIO spectra were used in the modelling}. Before being used for modeling, the original spectra were rebinned to a logarithmic wavelength grid with a radial velocity step of 3.5 km\,s$^{-1}$ in both spectral ranges.  Due to the small number of observations during epochs in which either of the systems was in an eclipse, no spin-axis orbital-plane misalignment for either of the components was assumed; thus, we set the parameter $\lambda$ = 0 deg.

The sum of synthetic spectra for all stellar components was fitted to the observed spectrum at each epoch.  Prior to summing, the synthetic spectra of individual components were convolved with a theoretical limb-darkened rotational profile. Mutual eclipses of the components were taken into account. Keplerian motion in both the inner and outer orbits was assumed. Synthetic spectra were computed using {\tt iSpec} \citep{2014A&A...569A.111B,2019MNRAS.486.2075B}, a program based on the {\tt SPECTRUM} code \citep{1994AJ....107..742G}. For the spectra, we assumed a solar metallicity ($\log [m/X] = 0.0$) and that 9\,000 $\leq$\, $T_{\rm eff}$ $\leq$\,12\,000 K. We also allowed for apsidal motion for the inner orbits. No other gravitational perturbations were assumed in this relatively simple model. We found that the principal atmospheric parameters corresponding to the best model for either of the spectral ranges are $\log g = 4.5$ for all four stars; $T_{\rm eff}$ = 11000 K for components 2, 3, and 4; and $T_{\rm eff}$ = 11\,500 K for component 1.\footnote{When describing our spectroscopic analysis, we refer to the stellar components of the A subsystem ($P = 2.94$\,d) as 1 and 2. The stellar components of the wider subsystem B ($P = 3.26$\,d) will be denoted as 3 and 4.}

We modeled the blue and red spectral ranges separately. In order to constrain the fit parameters---specifically the inclination angle $i$, orbital periods of inner binaries, $P_A$, $P_B$, and the apsidal motion rates of the binaries $P_{\rm aps}$---we included the \textit{TESS} photometry from sectors 7 and 34 in the modelling. The ellipsoidal light variations were modeled using the analytical approximations of \citet{Ellipsoidal}; these were found to be sufficiently accurate for both inner binaries.  While the semi-amplitudes of the RV changes $K_i$ are free parameters, the semi-amplitude of the radial-velocity variations of the center of masses of the binaries depend on the masses of the individual components ($M_i$, $i=1,2,3,4$), the inclination angle $i_{\rm out}$, eccentricity $e_{\rm out}$, and the orbital period $P_{\rm out}$ of the outer orbit. The semi-amplitudes of these radial velocity changes, $K_{\rm 12}$ and $K_{\rm 34}$, were hence computed using these parameters. 

The simultaneous modeling of the \textit{TESS} light curve and the spectra was performed via gradient-based optimization starting from multiple sets of trial parameters in order to arrive at the global minimum of $\chi^2$. Initial optimization runs showed that the inclination of the outer orbit is close to 90\degr~, but attempts to adjust it led to divergent results. Hence, the outer orbit was fixed at edge-on and the inclination was not adjusted. We were able to constrain the outer orbit and demonstrate that it is extremely tight, with $P_{\rm out} \sim 121$\,d. This differs significantly from the estimate provided by \citet{2021ARep...65..826V}, who cite a value over twenty times ours ($P_{\rm out}$ = 2420$\pm$40\,d or 6.62 years). In Section \ref{sec:discussion}, we discuss the differences in analyses, and lay out a case for why 121\,d is definitely the period of the outer orbit.

\begin{figure*}
    \centering
    \includegraphics[width=1.1\linewidth]{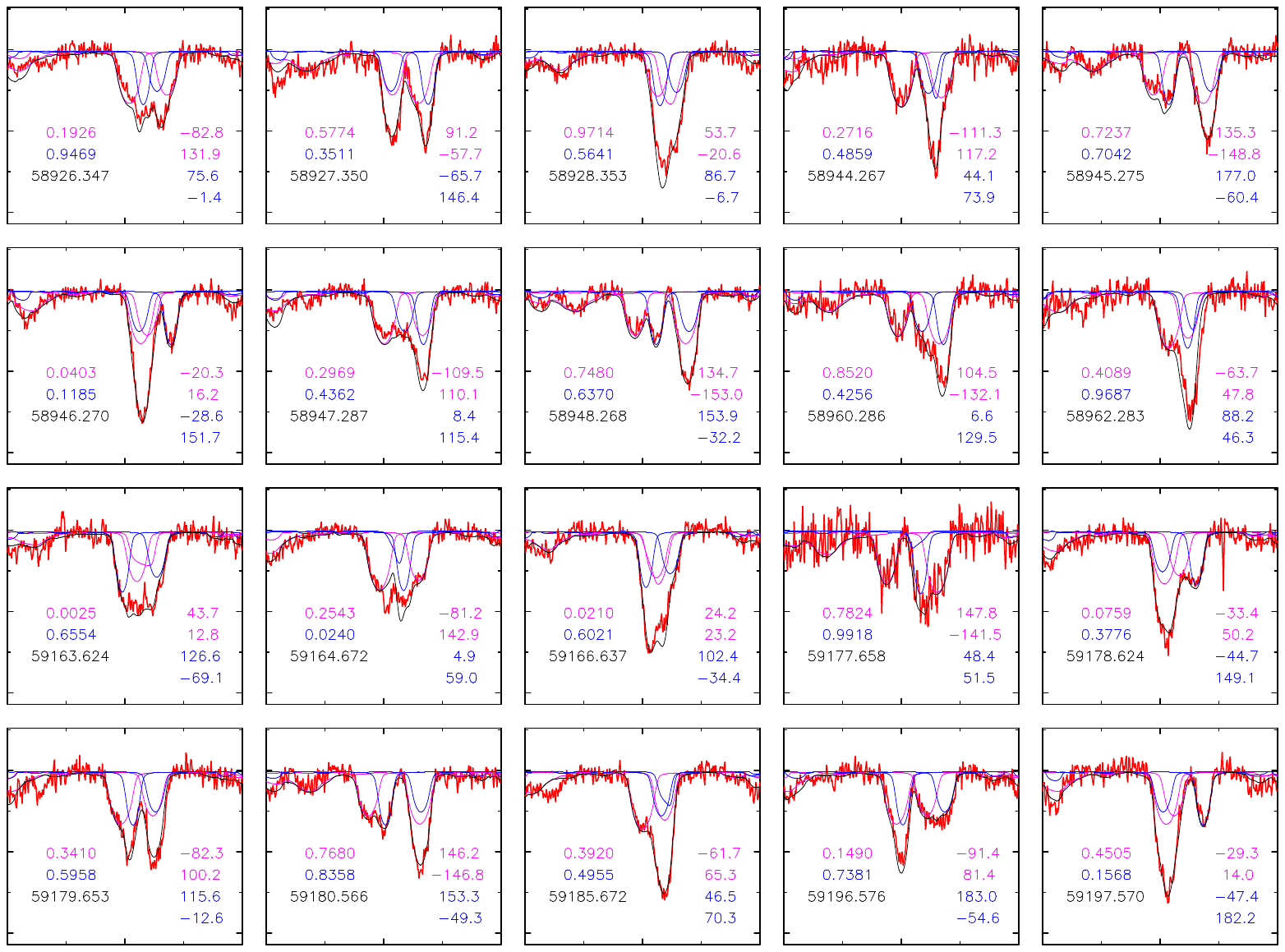}
    \caption{Spectrum of BU~CMi near the Mg {\sc ii} 4481 line (4470 - 4490 \AA~segment, plotted with red solid curve) and the optimum multi-component fit (black solid curve) for the first 20 spectra. Each panel shows the orbital phases of the inner subsystems and the epoch (HJD-2\,400\,000) typed in black on the left, and the model radial velocities of the binary subsystems on the right. Magenta is used for the predicted radial velocities, the orbital phase and spectrum of the A sub-system ($P = 2.94$\,d), while blue is used for the same plot elements for the B sub-system ($P=3.26$\,d). }
   \label{fig:SPspectra-directfit}
\end{figure*} % Fig. 4

%\ron{(Theo: is there any chance you could make the internal labels bold faced so they are easier to %read?)}

The parameters corresponding to the best models for the blue and red regions are listed in Table \ref{tbl:splc-model}, and the best-fit models to the first 20 observed spectra of BU~CMi in the blue wavelength range, close to the Mg {\sc ii} 4481 \AA~ line, are plotted in Figure \ref{fig:SPspectra-directfit}. The best-fit stellar parameters for the component stars are listed in Table~\ref{tbl:splc-absolute}.
For further analysis, we will equate the blue and yellow spectral ranges with the Johnson $B$ and $V$ passbands, and use $B$ and $V$ magnitudes where relevant.
\begin{table*}
\centering
\caption{Best-fit orbital elements of BU~CMi. These results are from the simultaneous modelling of the BU~CMi spectra and the \textit{TESS} light curves from sectors 7 and 34. 
The average of parameters determined from the blue spectral range (4380 - 4605.4 \AA, 4300 wavelength bins) and the orange spectral range, 4900 - 5379.6 \AA, 4380 - 4605.4 \AA, 8000 wavelength bins) are listed. The number of light curve data points for each solution is 2775. 
Fixed parameters that were not adjusted are marked by a superscript $f$; computed parameters are indicated by superscript $c$.}
\label{tbl:splc-model}
\begin{tabular}{lccccccc}
\hline
                                                              \multicolumn{8}{c}{orbital elements}                                                       \\
\hline
                    &            &   \multicolumn{2}{c}{subsystem A}  & \multicolumn{2}{c}{subsystem B}   & \multicolumn{2}{c}{subsystem A-B}\\
$P$                 & [days]     &  2.9406238      &   $\pm$0.0000014 &  3.2632996      &  $\pm$0.0000047 &  121.09         & $\pm$0.05  \\
$i$                 & [deg]      &    82.23        &   $\pm$0.07      &  84.355         &  $\pm$0.015     &   90$^f$        &    -        \\
$e$                 &            &    0.2087       &   $\pm$0.0003    &    0.2304       &  $\pm$0.0011    &   0.2633        & $\pm$0.0023 \\
$T_0$               & [BJD]      &   2458903.83783 &   $\pm$0.00027   & 2458900.3861    &  $\pm$0.0004    & 2459488.2       & $\pm$2.1    \\
$\omega$            & [deg]      &   300.05        &   $\pm$0.04      &   95.45         &  $\pm$0.05      &  146            & $\pm$5      \\
$P_{\rm apse}$      & [year]     &    24.91        &   $\pm$0.04      &   25.79         &  $\pm$0.08      &      -          &             \\
$K_{\rm pri}$       & [km/s]     &   126.3         &   $\pm$1.2       &   126.3         &  $\pm$0.3       &   49.06$^c$     &             \\
$K_{\rm sec}$       & [km/s]     &   134.5         &   $\pm$0.5       &   129.1         &  $\pm$0.2       &   48.51$^c$     &             \\
$\gamma$            & [km/s]     &       -         &        -         &      -          &        -        &   28.9          &  $\pm$0.5   \\
$a$                 & [R$_\odot$]&   14.97         &   $\pm$0.10      & 16.108          &  $\pm$0.016     &  225.3          &  $\pm$0.9   \\
\hline
\end{tabular}
\end{table*}

\begin{table}
\centering
\caption{Best-fit stellar parameters corresponding to the solutions listed in Table~\ref{tbl:splc-model} of BU~CMi. The averages of the parameter values resulting from the modelling of blue spectral range (4380 - 4605.4 \AA, 4300 wavelength bins) and orange spectral range (4900 - 5379.6 \AA, 8000 wavelength bins) are listed. The number of light curve data points is 2775. Fixed parameters are marked by a superscript $f$.}
\label{tbl:splc-absolute}
\begin{tabular}{lccccc}
\hline                                                          
                                &             &   \multicolumn{2}{c}{subsystem A}  & \multicolumn{2}{c}{subsystem B}  \\
\hline
$v \sin i_{\rm pri}$            & [km/s]      &    81.4         &   $\pm$0.7       &   63.1          &  $\pm$1.7      \\
$R_{\rm pri}/a$                 &             &     0.1560      &   $\pm$0.0015    &   0.1305        &  $\pm$0.0019   \\
$T_{\rm pri}$                   & [K]         & 11500$^f$       &     -            & 11000$^f$       &      -         \\
$M_{\rm pri}$                   & [M$_\odot$] &  2.68           &  $\pm$0.04       &  2.66           &  $\pm$0.02     \\ 
$R_{\rm pri}$                   & [R$_\odot$] &  2.33           &  $\pm$0.04       &  2.10           &  $\pm$0.03     \\ 
$L_{\rm pri}$                   & [L$_\odot$] & 86              &  $\pm$3          & 58.3            &  $\pm$1.7      \\ 
$\log g_{\rm pri}$              & [cgs]       &  4.130          &  $\pm$0.007      &  4.217          &  $\pm$0.012    \\ 
\hline
$v \sin i_{\rm sec}$            & [km/s]      &    71.6         &   $\pm$1.0       &   49.9          &  $\pm$0.8      \\
$R_{\rm sec}/a$                 &             &     0.1455      &   $\pm$0.0005    &   0.14643       &  $\pm$0.00016  \\
$T_{\rm sec}$                   & [K]         & 10850           &   $\pm$70        & 10400           &  $\pm$150      \\
$M_{\rm sec}$                   & [M$_\odot$] &  2.52           &  $\pm$0.06       &  2.605          &  $\pm$0.022    \\ 
$R_{\rm sec}$                   & [R$_\odot$] &  2.178          &  $\pm$0.023      &  2.358          &  $\pm$0.004    \\ 
$L_{\rm sec}$                   & [L$_\odot$] & 59.3            &  $\pm$0.5        &  59             &  $\pm$3        \\ 
$\log g_{\rm sec}$              & [cgs]       &  4.162          &  $\pm$0.003      &  4.108          &  $\pm$0.004    \\
\hline
\end{tabular}
\end{table}

\section{Spectra disentangling and atmospheric parameters}
\label{sec:spectral_disentanglement}

\begin{figure}
    \centering
    \includegraphics[width=1.0\linewidth]{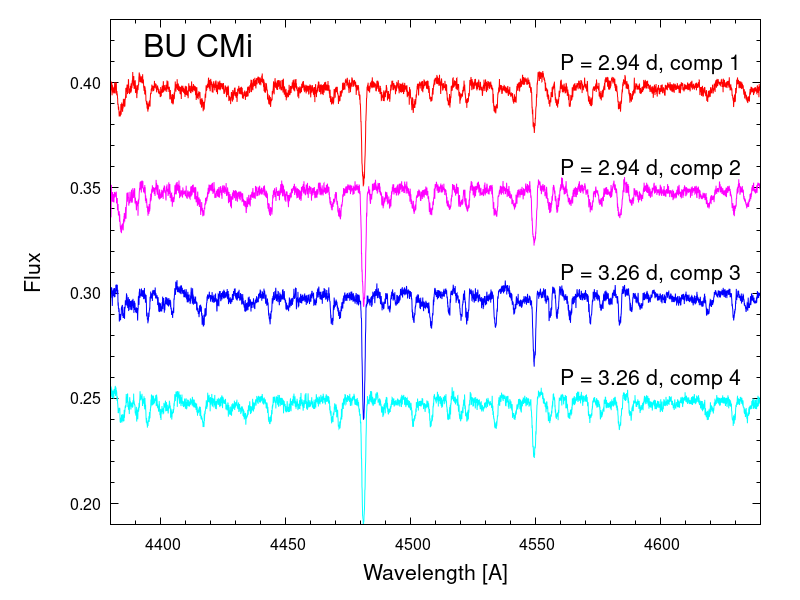}
    \caption{The individual disentangled component spectra of BU~CMi in the blue wavelength range. The spectra are shifted vertically by 0.05 in flux for clarity.}
   \label{fig:SPspectra-tomography}
\end{figure}

After obtaining a set of model radial velocities for both stellar components at all epochs of the spectroscopic observations, we then attempted to disentangle the composite spectrum in order to obtain individual spectra for each star (1, 2, 3, and 4). The iterative disentangling approach of \citet{1991ApJ...376..266B} was used, and the spectra were disentangled in both the blue and red ranges. As previously, a radial-velocity step size of 3.5 km\,s$^{-1}$ was used for both spectral ranges. The process converged to a solution for both spectral ranges, and the resulting composite spectra are plotted in Figure \ref{fig:SPspectra-tomography}. These spectra show small imperfections, perhaps from issues with the continuum rectification and artifacts from the disentangling.

However, our technique does not yield the flux ratio of the components \citep[see e.g.][]{2010ASPC..435..207P,2012IAUS..282..395L}. As a result, we corrected the component spectra for the contribution of the other components before proceeding. Properly normalizing the spectra will have an impact on the line depths and is possible using the component passband-specific luminosities derived from simultaneously modeling the spectra and the light curve.  Computed brightnesses for the $B$ and $V$ Johnson bands were assumed for the blue and red spectral ranges, respectively. As part of our modeling, we used the radii derived in Table \ref{tbl:splc-model} for the correction step and to determine the atmospheric parameters $T_{\rm eff}$, $\log g$, metallicity $\log [m/X]$.  The projected rotational velocities were taken from the results of the spectral fitting.  As before, we used the {\tt iSpec} code.

The temperatures and surface gravities for binary A are $T_{\rm pri}$ = 10760$\pm$360 K, $\log g_{\rm pri} = 4.23\pm0.29$, $T_{\rm sec}$ = 10180$\pm$430 K, $\log g_{\rm pri} = 4.11\pm0.30$. The corresponding parameters for binary B are $T_{\rm pri}$ = 10820$\pm$380 K, $\log g_{\rm pri} = 4.63\pm0.26$, $T_{\rm sec}$ = 10040$\pm$350 K, $\log g_{\rm pri} = 4.18\pm0.23$. The metallicity was fixed at the solar value. The component temperatures are slightly lower than those of the best-fitting templates. We also highlight our finding that component 4, in binary B ($P$=3.26\,d), is cooler than the remaining components. As part of our presentation of these results, we add the caveat that the atmospheric parameters strongly depend on the technique used to normalize the individual disentangled spectra. Using different flux ratios would change the line depth and result in different temperatures and/or metallicities for each of the components in the system.

 % Fig. 2

\section{Spectro-photodynamical Analysis}
\label{sect:photodynamics}
Independent of the spectroscopic analyses described in Sect.~\ref{sec:Theo_analyses}, we carried out a simultaneous, joint analysis of the \textit{TESS} light curve, the ETV curves calculated from the mid-times of all the observed eclipses (i.e., from \textit{TESS} and ground-based measurements; see Sect.~\ref{sec:ToMs}), radial velocities (determined by Gaussian-profile fitting to the BFs; see Sect.~\ref{sec:RVs}), as well as all the available multi-wavelength SED data for BU CMi using the {\sc Lightcurvefactory} software package \citep{borkovitsetal19,borkovitsetal20}.  During the analysis we followed the same steps as for TIC~454140642 and TIC~219006972 (\citealt{kostovetal21}; \citealt{kostovetal23}), two other compact quadruple systems discovered with \textit{TESS}. For further details of the photodynamical analysis, we refer the reader to Sect.~5.1 of \citet{kostovetal21}. Here, we describe only the steps in the data preparation that are specific to this system.

In order to retain similar sampling and to reduce computational costs, we binned the 2-minute cadence datasets from Sectors 34 and 61 to 1800 sec, similar to the sector 7 FFI-cadence light curve. When comparing some preliminary fits to the short- and long-cadence light curves, we found no statistically significant discrepancies in the resultant parameters and, hence, we decided to use exclusively the 1800-sec cadence light curves for all three available \textit{TESS} sectors.

During our earlier initial analysis runs (made in 2021 and 2022), we also made use of a second set of light curves compiled from the ground-based photometric follow up observations. However, when the sector 61 observations of the \textit{TESS} spacecraft became available and, hence, the high-quality and homogeneous satellite photometry then covered more than a 4-year-long interval, we decided to no longer use the more inhomogeneous and less accurate ground-based photometric data for direct light curve fitting. Naturally, the eclipse times derived from these ground-based observations did continue to be used for the ETV analysis part of the simultaneous, joint spectro-photodynamical fitting process.

Because our software package {\sc Lightcurvefactory} is unable to handle direct fitting of the spectral lines (see Sect.~\ref{sec:Theo_analyses}) and fits only direct RV data, we used a second, different approach. We mined out model-independent RV data from all available spectra in the manner described briefly in Sect.~\ref{sec:KonkolyRozhen}. Due to the highly blended spectral lines, the correct identifications of the individual stellar components were quite problematic in the case of several spectra.  Hence, after obtaining the first model fits, we had to change the labeling of the stars in the RV data in several cases.  The RV data used for the spectro-photodynamical analysis are listed in Table~\ref{tbl:rvs}.

In addition to the \textit{TESS} light curves and the RV data, as was mentioned before, we used the four ETV curves (primary and secondary ETV data for both binaries; see Tables~\ref{Tab:BU_CMi_A_(TIC_271204362A)_ToM} and \ref{Tab:BU_CMi_B_(TIC_271204362B)_ToM}).

Finally, we also used the catalog passband magnitudes listed in Table~\ref{tab:EBparameters} for the SED analysis. For this analysis, similar to our previous works, we used a minimum uncertainty of $0.03$ mag for most of the observed passband magnitudes.  This was done in order to avoid an outsized contribution from the extremely precise Gaia magnitudes, as well as to counterbalance the uncertainties inherent in our interpolation method during the calculations of theoretical passband magnitudes that are part of the fitting process. The only exception is the WISE $W4$ magnitude, for which the uncertainty was set to 0.3 mag.

\begin{figure*}
    \centering
    \includegraphics[width=0.48\linewidth]{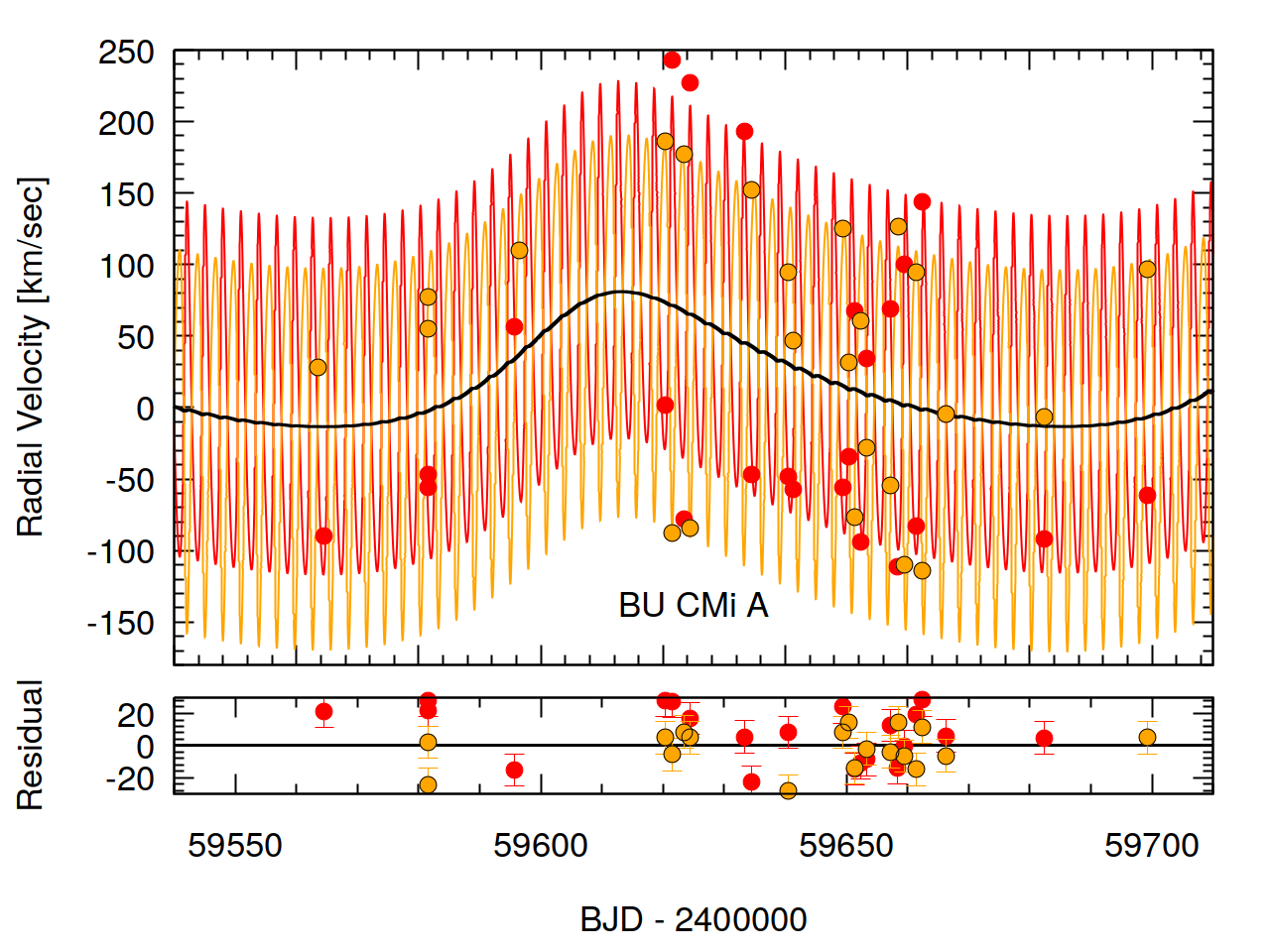}\includegraphics[width=0.48\linewidth]{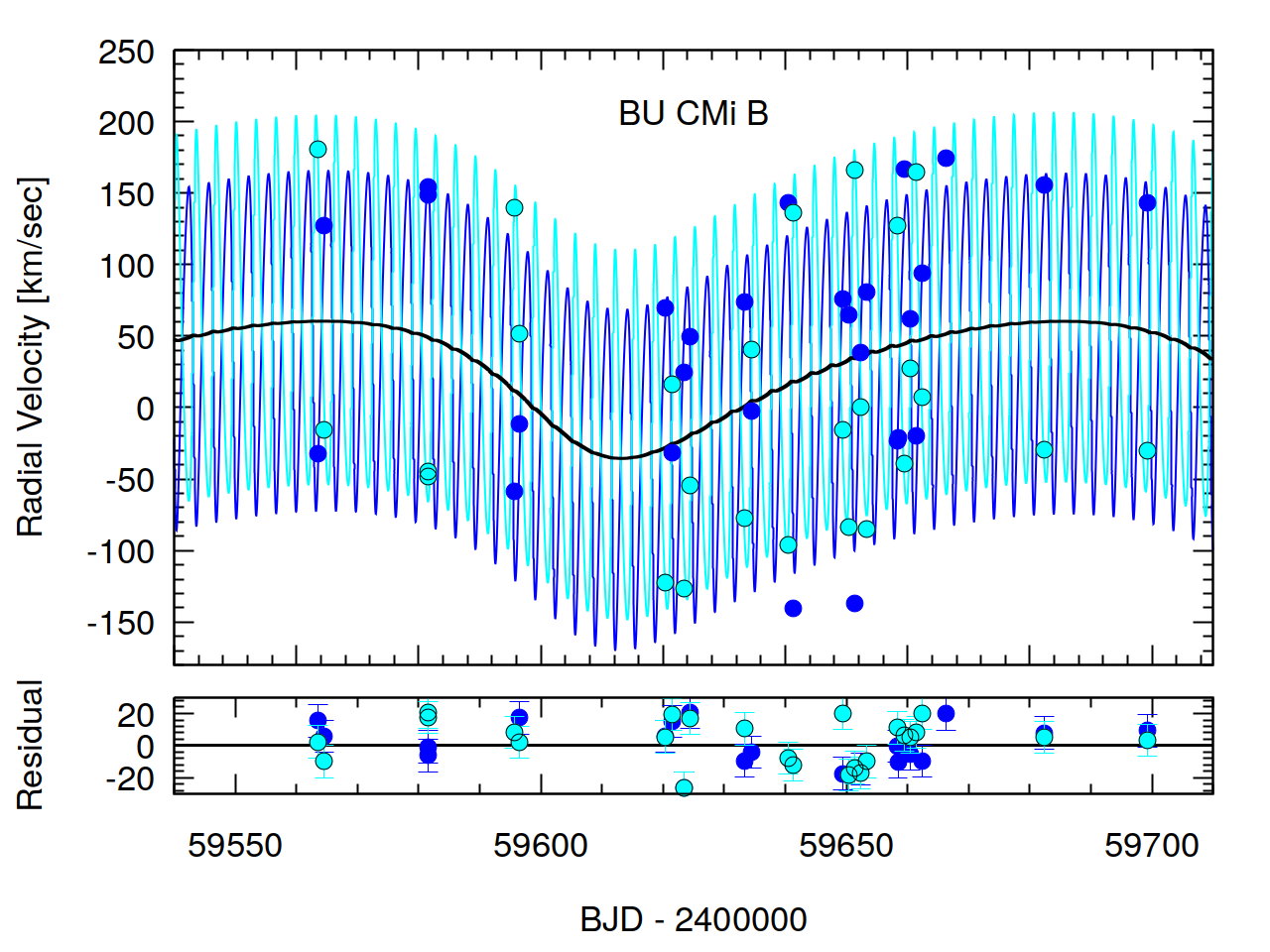}
    \includegraphics[width=0.48\linewidth]{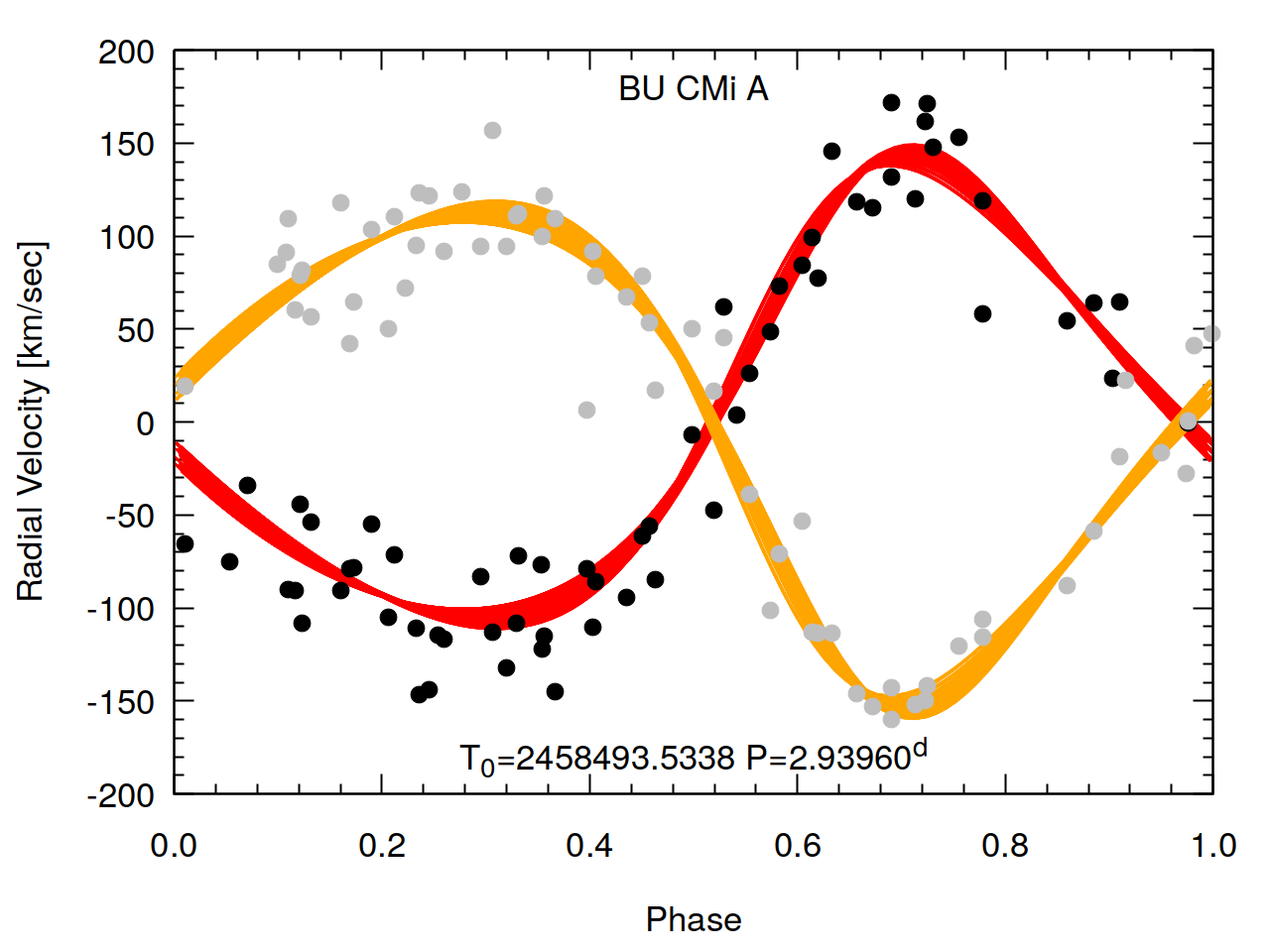}\includegraphics[width=0.48\linewidth]{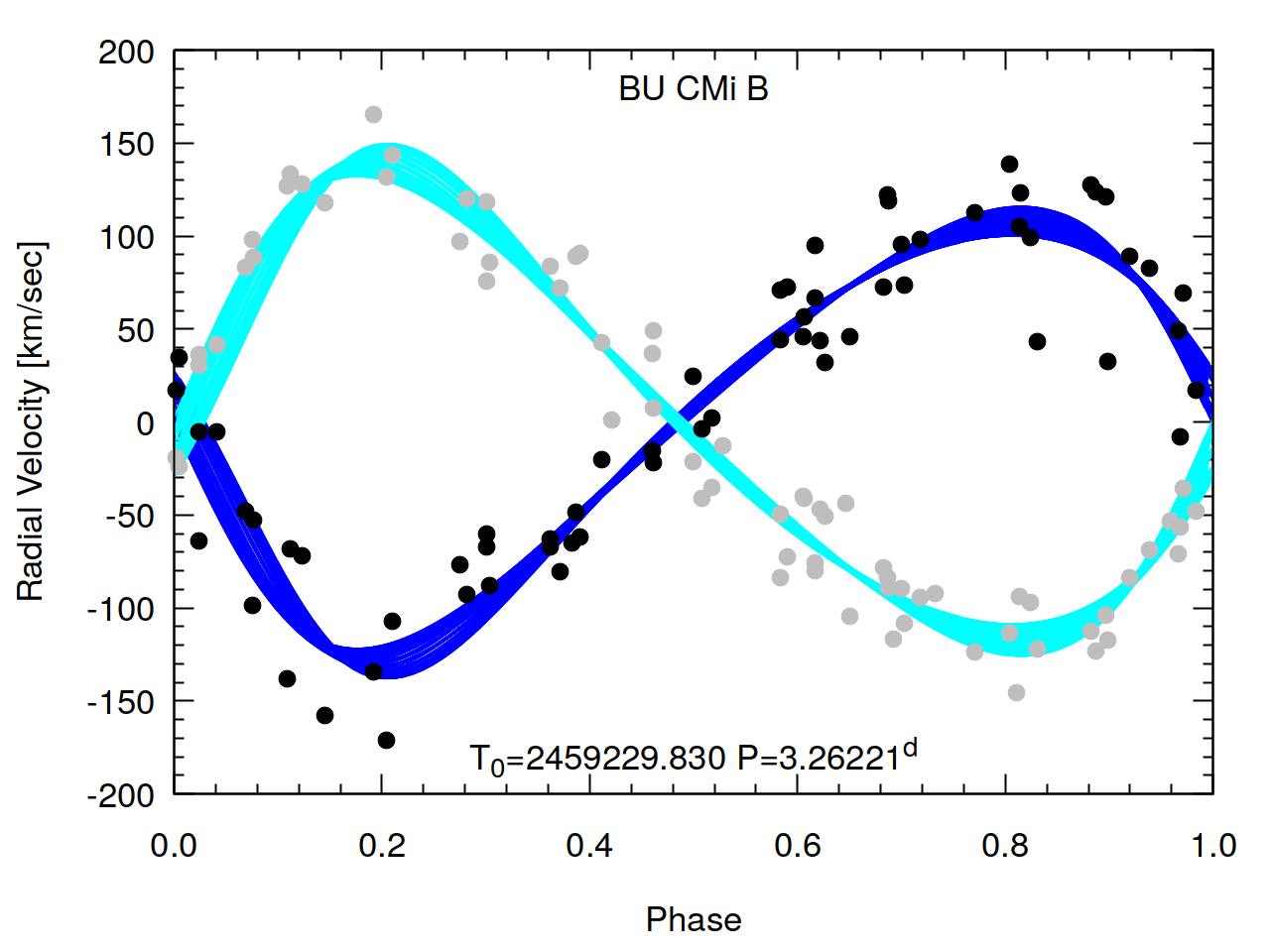}
    \caption{{\it Upper panels:} A 170-day interval of the radial velocity data of binaries A and B (left and right, respectively), together with the best-fit spectro-photodynamical model. Red and blue dots and curves refer to the primary components ($Aa$, $Ba$, respectively), while orange and cyan symbols stand for the secondaries ($Ab$, $Bb$). The black curves represent the varying radial velocities of the center of masses of the binaries along their revolution around each other. {\it Lower panels:} Phase-folded RV curves of both binaries for the full observational intervals, after the removal of the model-derived motions of the centres of mass of both binaries. The varying shape (thickness) of the model curves is the consequence of the dynamical perturbations (primarily the apsidal motion). Black and gray circles represent the observed data corrected for the orbital motion of the two binaries around each other.}
   \label{fig:RV}
\end{figure*}

Table~\ref{tbl:simlightcurve} lists the median values of the stellar and orbital parameters of the BU CMi quadruple system that have been either adjusted, internally constrained, or derived from the MCMC posteriors, together with the corresponding $1\sigma$ statistical uncertainties. Sections of the light curves, the ETV and RV curves of the lowest $\chi^2_\mathrm{global}$ solution are plotted in Figs.~\ref{fig:TESSlightcurves+model},  \ref{fig:etv}, and \ref{fig:RV}, respectively.

One caveat should be noted regarding the proper interpretation of the orbital parameters listed in Table~\ref{tbl:simlightcurve}. The tabulated orbital elements, with a few exceptions, are so-called `instantaneous osculating elements,' which are valid at the moment of the cited epoch $t_0$. Thus, they cannot be simply compared with those orbital elements that are deduced directly from photometric and/or spectroscopic observations. These latter orbital elements can be considered to be some kind of long-term averaged orbital elements and connected to such Keplerian orbits that represent the approximations of the time-averaged envelopes of the true (non-Keplerian) motions. This question was discussed in more detail in Sect.~5.1 of \citet{kostovetal21}. 

Regarding the exceptions, in the (second) row $\overline{P}_\mathrm{obs}$ we give the average or, `observable', periods for the three orbits (A, B, AB) which were obtained from a longer-term numerical integration initiated with the parameters of the best-fit spectro-photodynamical model. The values given for the inner binaries (A, B) stand for the long-term average of their eclipsing periods (which technically means that calculating the ETV curves with these periods, the averages of the ETVs over full apsidal motion cycles remain constant). In the case of the outer orbit, however, $\overline{P}_\mathrm{obs}$ was determined as the time average of consecutive periastron passages, i.e., this is the average anomalistic period of the outer orbit.

The other exception is the first row of the apsidal motion related parameters, $P_\mathrm{apse}^\mathrm{obs}$. Here the duration of an apsidal motion cycle, i.e., the time needed for the complete, $360\degr$ variation of the observable arguments of periastron of the three orbits, $\omega_\mathrm{A,B,AB}$, are given. The other apsidal motion parameters are calculated internally by the software package {\sc Lightcurvefactory} with the use of the usual analytic formulae \citep[see][Sect. 6.2, for details]{kostovetal21}.

 % Fig. 2

%The orbital configuration of the system, as seen from above the plane of the outer orbit, is shown in Figure \ref{fig:top_view} for a few outer periods. 

\begin{table*}
\centering
\caption{Median values of the parameters of BU CMi from the double EB simultaneous light curve, $2\times$ SB2 radial velocity, double ETV, joint SED and \texttt{PARSEC} 
evolutionary tracks solution from {\sc Lightcurvefactory.}}
\begin{tabular}{lccccc}
\hline
\multicolumn{6}{c}{Orbital elements$^a$} \\
\hline
   & \multicolumn{3}{c}{Subsystem}  \\
   & \multicolumn{2}{c}{A} & \multicolumn{2}{c}{B} & A--B \\
  \hline
$P_\mathrm{a}$ [days]            & \multicolumn{2}{c}{$2.937533_{-0.000089}^{+0.000096}$} & \multicolumn{2}{c}{$3.255815_{-0.000142}^{+0.000230}$}   & $121.793_{-0.035}^{+0.027}$ \\
$\overline{P}_\mathrm{obs}^b$ [days]          & \multicolumn{2}{c}{$2.93959_{-0.00001}^{+0.00001}$} & \multicolumn{2}{c}{$3.26222_{-0.00001}^{+0.00001}$}   & $121.50_{-0.02}^{+0.02}$ \\
semimajor axis  [$R_\odot$]      & \multicolumn{2}{c}{$14.68_{-0.17}^{+0.30}$}            & \multicolumn{2}{c}{$15.64_{-0.19}^{+0.30}$}              & $220.9_{-2.6}^{+4.5}$ \\  
$i$ [deg]                        & \multicolumn{2}{c}{$83.35_{-0.12}^{+0.13}$}            & \multicolumn{2}{c}{$83.92_{-0.13}^{+0.14}$}              & $83.79_{-0.25}^{+0.26}$  \\
$e$                              & \multicolumn{2}{c}{$0.2191_{-0.0007}^{+0.0008}$}       & \multicolumn{2}{c}{$0.2257_{-0.0009}^{+0.0009}$}         & $0.2728_{-0.0034}^{+0.0040}$ \\  
$\omega$ [deg]                   & \multicolumn{2}{c}{$101.41_{-0.14}^{+0.11}$}           & \multicolumn{2}{c}{$257.52_{-0.12}^{+0.18}$}             & $141.0_{-2.9}^{+4.3}$ \\
$\tau^c$ [BJD]     & \multicolumn{2}{c}{$2\,458\,445.1142_{-0.0019}^{+0.0023}$}           & \multicolumn{2}{c}{$2\,458\,492.4461_{-0.0011}^{+0.0018}$} & $2\,458\,513.59_{-0.57}^{+0.93}$\\
$\Omega$ [deg]                   & \multicolumn{2}{c}{$0.14_{-0.15}^{+0.15}$}                              & \multicolumn{2}{c}{$0.0$}               & $0.09_{-0.17}^{+0.17}$ \\
$(i_\mathrm{m})_{A-...}$ [deg]   & \multicolumn{2}{c}{$0.0$}                            & \multicolumn{2}{c}{$0.61_{-0.14}^{+0.15}$}                 & $0.45_{-0.17}^{+0.24}$ \\
$(i_\mathrm{m})_{B-...}$ [deg]   & \multicolumn{2}{c}{$0.61_{-0.14}^{+0.15}$}           & \multicolumn{2}{c}{$0.0$}                                  & $0.26_{-0.10}^{+0.17}$ \\
%$\omega^\mathrm{dyn}$ [deg      ]& \multicolumn{2}{c}{$~$} & \multicolumn{2}{c}{$_{-}^{+}$} & $_{-}^{+}$ \\
%$i^\mathrm{dyn}$ [deg] & \multicolumn{2}{c}{$0.31_{-0.17}^{+0.28}$} & \multicolumn{2}{c}{$0.41_{-0.21}^{+0.27}$} & $0.06_{-0.04}^{+0.06}$ \\
%$\Omega^\mathrm{dyn}$ [deg] & \multicolumn{2}{c}{$_{-}^{+}$} & \multicolumn{2}{c}{$_{-}^{+}$} & $_{-}^{+}$ \\
%$i_\mathrm{inv}$ [deg] & \multicolumn{5}{c}{$87.67_{-0.40}^{+0.38}$} \\
%$\Omega_\mathrm{inv}$ [deg] & \multicolumn{5}{c}{$-0.13_{-0.24}^{+0.24}$} \\
\hline
mass ratio $[q=m_\mathrm{sec}/m_\mathrm{pri}]$ & \multicolumn{2}{c}{$0.939_{-0.012}^{+0.011}$} & \multicolumn{2}{c}{$0.917_{-0.016}^{+0.015}$} & $0.982_{-0.005}^{+0.006}$ \\
$K_\mathrm{pri}$ [km\,s$^{-1}$] & \multicolumn{2}{c}{$124.86_{-1.76}^{+2.47}$} & \multicolumn{2}{c}{$118.91_{-2.02}^{+2.12}$} & $47.09_{-0.63}^{+0.78}$ \\ 
$K_\mathrm{sec}$ [km\,s$^{-1}$] & \multicolumn{2}{c}{$132.84_{-1.76}^{+2.78}$} & \multicolumn{2}{c}{$129.76_{-2.23}^{+2.37}$} & $47.87_{-0.58}^{+1.02}$ \\ 
$\gamma$ [km/s]                 & \multicolumn{4}{c}{$-$} & $23.22_{-0.07}^{+0.07}$\\ 
  \hline
  \multicolumn{6}{c}{Apsidal motion related parameters$^d$} \\
  \hline
$(P_\mathrm{apse}^\mathrm{obs})^b$ [year] &  \multicolumn{2}{c}{$28.7_{-0.1}^{+0.1}$} & \multicolumn{2}{c}{$25.1_{-0.1}^{+0.1}$} & $145.3_{-0.2}^{+0.2}$\\ $P_\mathrm{apse}$ [year] & \multicolumn{2}{c}{$31.75_{-0.13}^{+0.13}$} & \multicolumn{2}{c}{$28.96_{-0.10}^{+0.09}$} & $151.6_{-0.8}^{+0.7}$\\ 
$P_\mathrm{apse}^\mathrm{dyn}$ [year] & \multicolumn{2}{c}{$14.09_{-0.14}^{+0.71}$} & \multicolumn{2}{c}{$12.84_{-0.36}^{+1.15}$} & $21.7_{-0.3}^{+1.8}$\\ 
%$P_\mathrm{node}^\mathrm{dyn}$ [year] & \multicolumn{3}{c}{$24.99_{-0.08}^{+0.10}$}  & \multicolumn{3}{c}{$225_{-3}^{+3}$} \\
$\Delta\omega_\mathrm{3b}$ [arcsec/cycle] & \multicolumn{2}{c}{$718_{-35}^{+8}$} & \multicolumn{2}{c}{$885_{-74}^{+26}$} & $19\,912_{-1\,490}^{+309}$ \\ 
$\Delta\omega_\mathrm{GR}$ [arcsec/cycle] & \multicolumn{2}{c}{$2.907_{-0.067}^{+0.122}$} & \multicolumn{2}{c}{$2.694_{-0.066}^{+0.104}$} & $0.394_{-0.009}^{+0.017}$ \\ 
$\Delta\omega_\mathrm{tide}$ [arcsec/cycle] & \multicolumn{2}{c}{$18.73_{-0.37}^{+0.38}$} & \multicolumn{2}{c}{$12.67_{-0.27}^{+0.26}$} & ... \\  
  \hline  
\multicolumn{6}{c}{Stellar parameters} \\
\hline
   & Aa & Ab &  Ba & Bb & \\
  \hline
 \multicolumn{6}{c}{Relative quantities} \\
  \hline
fractional radius [$R/a$]               & $0.1528_{-0.0016}^{+0.0017}$ & $0.1409_{-0.0016}^{+0.0013}$  & $0.1422_{-0.0014}^{+0.0018}$ & $0.1278_{-0.0020}^{+0.0019}$ & \\
relative temperature [$T_\mathrm{x}/T_\mathrm{Ba}$] & $1.033_{-0.007}^{+0.007}$ & $0.971_{-0.004}^{+0.003}$ & $1$ & $0.955_{-0.010}^{+0.009}$ \\
fractional flux [in \textit{TESS}-band] & $0.2781_{-0.0090}^{+0.0105}$ & $0.2229_{-0.0045}^{+0.0037}$  & $0.2697_{-0.0060}^{+0.0074}$ & $0.2011_{-0.0093}^{+0.0089}$ & \\
%fractional flux [in \textit{SWASP}-band]& $0.311_{-0.008}^{+0.010}$    & $0.245_{-0.009}^{+0.008}$     & $0.246_{-0.010}^{+0.007}$    & $0.200_{-0.009}^{+0.007}$    & \\
%fractional flux [in $I_C$-band]         & $0.288_{-0.002}^{+0.002}$    & $0.213_{-0.003}^{+0.003}$     & $0.252_{-0.005}^{+0.001}$    & $0.249_{-0.007}^{+0.002}$    & \\
 \hline
 \multicolumn{6}{c}{Physical Quantities} \\
  \hline 
 $m$ [M$_\odot$]   & $2.536_{-0.089}^{+0.160}$ & $2.384_{-0.087}^{+0.143}$ & $2.521_{-0.090}^{+0.152}$ & $2.320_{-0.096}^{+0.125}$ & \\
 $R^c$ [R$_\odot$] & $2.249_{-0.038}^{+0.046}$ & $2.072_{-0.042}^{+0.045}$ & $2.235_{-0.042}^{+0.034}$ & $2.003_{-0.048}^{+0.043}$ & \\
 $T_\mathrm{eff}^e$ [K]& $10\,570_{-367}^{+455}$ & $10\,283_{-410}^{+381}$ & $10\,593_{-433}^{+380}$   & $10\,062_{-333}^{+428}$   & \\
 $L_\mathrm{bol}^e$ [L$_\odot$] & $55.60_{-7.83}^{+14.08}$ & $42.60_{-6.90}^{+8.89}$ & $57.01_{-10.79}^{+9.69}$ & $36.33_{-4.98}^{+9.12}$ &\\
 $M_\mathrm{bol}^e$ & $0.41_{-0.24}^{+0.16}$    & $0.70_{-0.21}^{+0.19}$    & $0.38_{-0.17}^{+0.23}$    & $0.87_{-0.24}^{+0.16}$    &\\
% $M_V^b           $ & $0.91_{-0.01}^{+0.01}$    & $1.26_{-0.02}^{+0.01}$    & $1.07_{-0.02}^{+0.02}$    & $1.08_{-0.02}^{+0.04}$    &\\
 $\log g^e$ [dex]   & $4.139_{-0.010}^{+0.012}$ & $4.184_{-0.007}^{+0.008}$ & $4.146_{-0.013}^{+0.010}$ & $4.202_{-0.011}^{+0.009}$ &\\
 \hline
\multicolumn{6}{c}{Global Quantities} \\
\hline
$\log$(age) [dex] &\multicolumn{5}{c}{$8.422_{-0.081}^{+0.077}$} \\
$ [M/H]$  [dex]      &\multicolumn{5}{c}{$0.023_{-0.100}^{+0.080}$} \\
$E(B-V)$ [mag]    &\multicolumn{5}{c}{$0.051_{-0.018}^{+0.020}$} \\
$(M_V)_\mathrm{tot}^e$           &\multicolumn{5}{c}{$-0.61_{-0.14}^{+0.10}$} \\
distance [pc]                &\multicolumn{5}{c}{$243_{-6}^{+10}$}  \\  
\hline
\end{tabular}
\label{tbl:simlightcurve}

{\em Notes.} (a) Instantaneous, osculating orbital elements at epoch $t_0=2\,458491.5$;  (b) Periods derived numerically from the best-fit photodynamical solutions (see text for details); (c) Time of periastron passage; (d) Apart from $P_\mathrm{apse}^\mathrm{obs}$, the other apsidal motion parameters were calculated analytically in the manner described in \citet{kostovetal21}; (e) Interpolated from the \texttt{PARSEC} isochrones.
\end{table*}

\section{Discussion and Implications}
\label{sec:discussion}

\subsection{System parameters}

Our two independent analyses (Sects.~\ref{sec:Theo_analyses}-\ref{sect:photodynamics}) have resulted in very similar results, at least qualitatively; however, quantitatively, a number of the discrepancies exceed the estimated uncertainties. For example, both models agree that BU CMi consists of four very similar hot stars. Moreover, the mass ratio of the two primaries (i.e., $m_\mathrm{Ba}/m_\mathrm{Aa}$) are $\sim$0.993 and $\sim$0.992, according to the direct spectral fitting and the spectro-photodynamical solutions, respectively (i.e., they agree well within 0.1\%). The direct spectral fitting approach, however, resulted in systematically more massive (by $\sim$4-5\%) and hotter (by $\sim$3-8\%) components than the spectro-photodynamical analysis. At the moment we are unable to resolve completely these discrepancies. One should keep in mind, however, that the individual masses are primarily controlled by the RV data which, in the current situation, are quite uncertain for the reasons discussed in Sects.~\ref{sec:RVs} and \ref{sec:Theo_analyses}. Regarding the temperatures, our attempts to obtain reliable spectroscopic solutions with the use of slightly lower temperature templates have failed. On the other hand, if one compares the photometric distance calculated from the spectro-photodynamical solution $d_\mathrm{phot}=245\pm8$\,pc with the Gaia DR3-derived one $d_\mathrm{GaiaDR3}=247\pm2$\,pc \citep{2021AJ....161..147B} the agreement looks perfect. In contrast to this, in the case of the direct spectral-fitting solution which gives hotter and more massive (and, hence, larger) stellar components, the total luminosity of the quadruple was found to be larger by $\sim$66\% (compare the derived individual luminosities in Tables~\ref{tbl:splc-absolute} and \ref{tbl:simlightcurve}) resulting in a $\sim29\%$ larger photometric distance. However, one should keep in mind that in the Gaia DR3 astrometric solutions, stellar multiplicity was not taken into account and hence the Gaia-derived distance might be subject to some systematic errors. Thus, in conclusion, we believe that we are not currently in a position to prefer one or the other solution; hence, we conclude that the true uncertainties of our solutions, at least in the masses and temperature, must be around 5-7\%. The quantitatively similar results of the two independent approaches clearly show that the parameters are robust and BU CMi is indeed a very tight quadruple with $\sim$120 day outer orbital period. The differences in the orbital and component parameters between the two analyses give us, however, an independent check of the parameter uncertainties. The similar brightness of the inner binaries makes the astrometric signal of the system's photocenter tiny; thus we cannot expect that Gaia would provide independent orbital parameters. Because the photodynamical analysis is more complex and in addition to RVs it takes into account all available photometric data (including observed colors and brightness), we prefer the corresponding parameters.

Turning to the orbital geometry of the system, we find that BU CMi is not only the most compact but also an extremely flat, 2+2 type quadruple system. All three orbital planes (that of the two inner EBs and also the outer orbital plane) are well aligned within 1\degr. Note, this is a common feature of the previous two \textit{TESS}-discovered compact doubly eclipsing 2+2 type quadruple systems: TIC~454140642 \citep{kostovetal21} and TIC~219006972 \citep{kostovetal23}. In drawing any general conclusions from the common flatness of these new compact 2+2 quadruples, however, one should keep in mind that these findings might be biased by observational selection effects, at least, in part. This is so because, in the case of substantially inclined orbital planes, the three planes would precess with different amplitudes and periods on timescales of decades or centuries (depending on the mutual inclination angles), and this would make it very unlikely that both inner binaries would show eclipses at the same time. And, in the absence of eclipses in either or both of the binaries, there would be no (or only a very minor) chance of detecting the given target as a compact 2+2 system. On the other hand, however, such a selection effect does not explain the {\it extreme} flatness (i.e. orbital alignments within $1-2\degr$). For example, the current binaries A and B would already produce eclipses for inclinations $i_\mathrm{A}\gtrsim72\fdg9$ and $i_\mathrm{B}\gtrsim74\fdg3$, from which it follows that, keeping the inclination of the system's invariable plane at the currently derived value of $i_0=83\fdg8\pm0\fdg2$, both inner binaries would continuously produce well detectable eclipses in the case of mutual inclinations of, let's say, $i_\mathrm{A-AB;B-AB}=5\degr$. So, in conclusion, in our view the very strong flatness is likely a consequence of the formation processes of the most compact 2+2 systems, but, due to the strong observational bias described above, this conclusion is not highly robust.

Besides the similarities of BU CMi to the previously mentioned two quadruples, there are strong differences as well. For example, in the case of the other two quadruples, the inner binaries are in nearly circular orbits, while the inner binaries of BU CMi have remarkably substantial eccentricities ($e_\mathrm{A}=0.2191\pm0.0008$ and $e_\mathrm{B}=0.2257\pm0.0009$, respectively). As BU CMi has the smallest $P_\mathrm{A,B}/P_\mathrm{out}$ ratio of these three quadruples and, moreover, contains the closest inner binaries amongst these three systems, these findings at first sight appear to be quite unexpected for a number of reasons. First, the less tight a hierarchical system is, the less effective are the gravitational perturbations of the outer component(s) acting upon the Keplerian motion(s) of the inner pair(s). Second, due to the compactness of the inner binaries, the fractional radii of the constituent stars in BU CMi exceed the same quantities of the stars in the other two quadruples by factors of $4-5$. Hence, one can expect that tidal forces and tidal dissipation, which are related to the fifth (for equilibrium tides) and eighth order (for tidal dissipation) of the fractional radii are much more effective in BU CMi than in the other two systems.

These discrepancies can be resolved by considering the facts that (i) in contrast to TICs~454140642 and 219006972, BU CMi is formed by four early-type, massive, radiative stars, in which tidal dissipation is much less effective; and moreover, (ii) this system is much younger (i.e., $\sim300$\,Myr-old, in contrast to the other two quadruples which are several Gyr old). Thus, one may conclude that there has been insufficient time for the circularizaton of the inner orbits since the formation of the quadruple system BU CMi. 

\begin{figure*}
    \centering
    \includegraphics[width=0.48\linewidth]{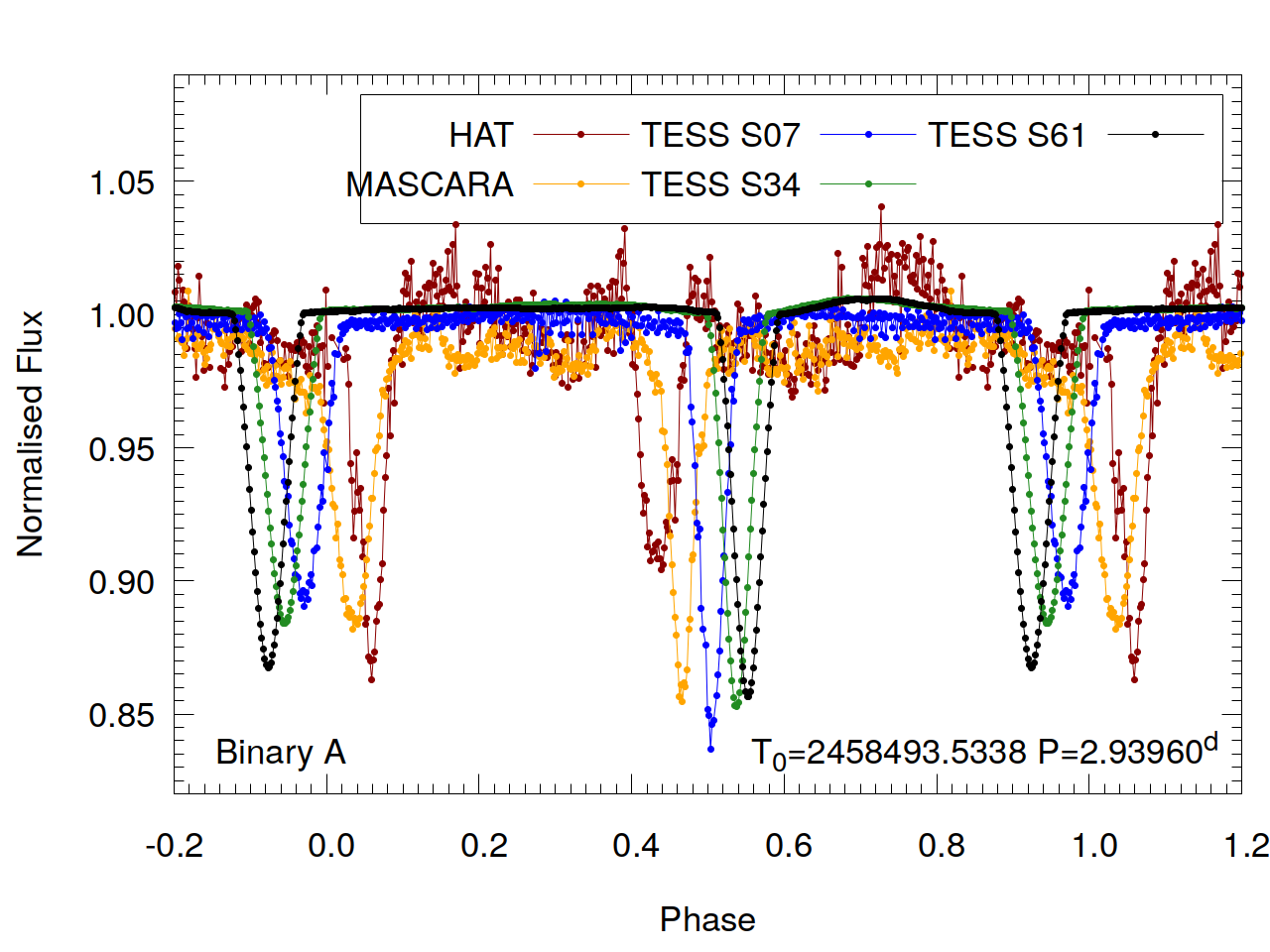}\includegraphics[width=0.48\linewidth]{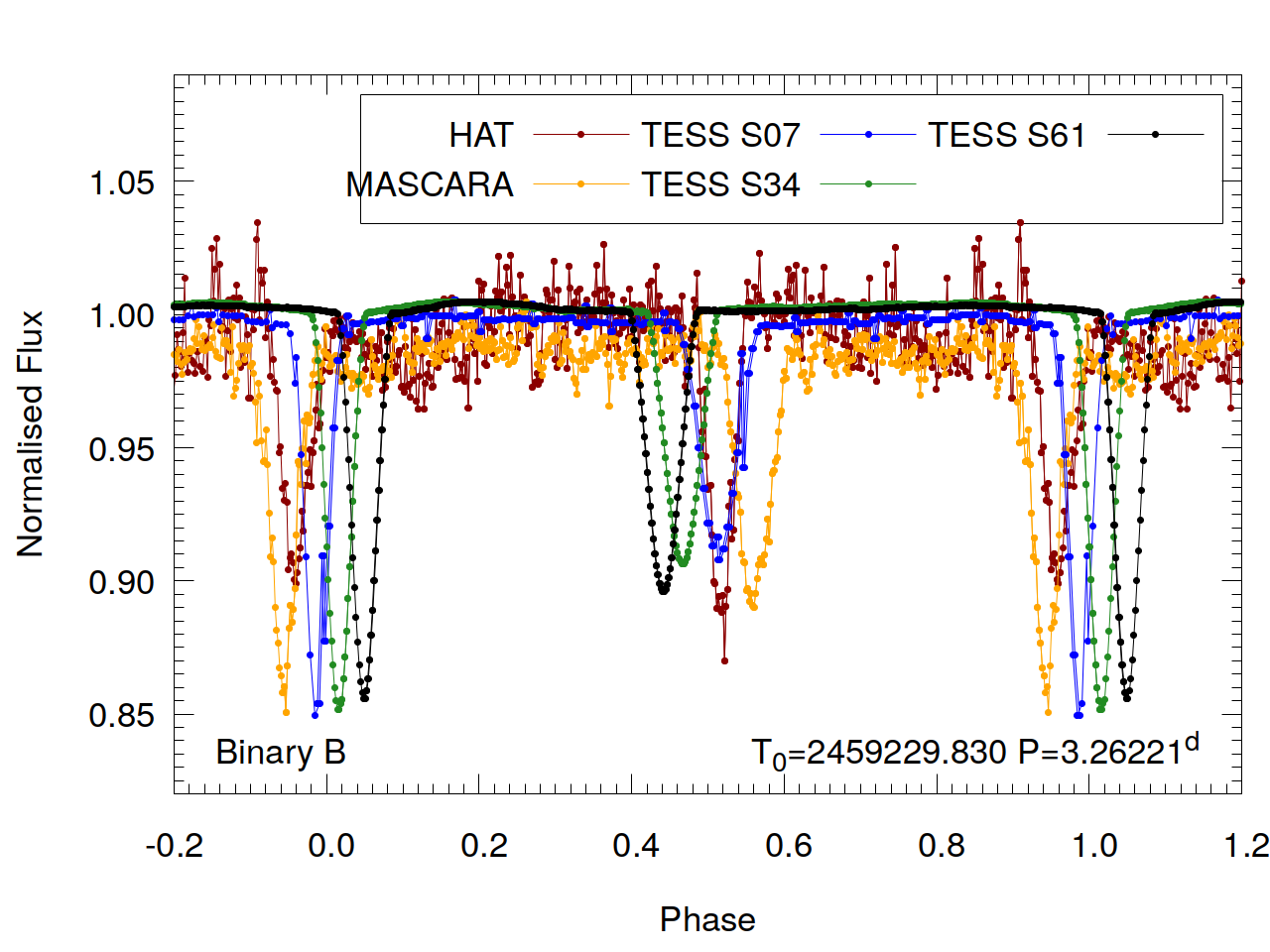}
    \caption{Illustration of the reversal of the primary and secondary eclipse depths of both binaries during an apsidal cycle. Dark red, orange, blue and green symbols stand for the folded, binned, averaged light curves of binaries A and B (left and right, respectively), calculated from the \textit{HAT} (red), \textit{MASCARA} (orange), and \textit{TESS} sectors 7 (blue), 34 (green) and 61 (black) observations. As one can see, the primary/secondary eclipse depth ratios of both binaries have reversed in between the oldest \textit{HAT} and the newer observations.}
   \label{fig:ABfolds}
\end{figure*}

The significant inner eccentricities in BU CMi, as well as the non-edge-on view of the two inner orbital planes ($i_\mathrm{A}=83\fdg4\pm0\fdg1$; $i_\mathrm{B}=83\fdg9\pm0\fdg1$), together with the very similar surface brightnesses of the binary stars, have an interesting observational consequence. Namely, depending on the orientations of the orbital ellipses relative to the Earth, the depths of the two consecutive eclipses in each EB alter each others. Regarding binary A, currently the periastron passage, i.e., the smallest separation between the two stars, is much closer to the inferior conjunction of the more massive (primary) component.  Hence, due to the larger fraction of the occulted stellar disk of the secondary, the deeper light minimum occurs during this event, i.e., when the more massive, hotter star eclipses the less massive and cooler component. In contrast to this, during the \textit{HAT} observations, nearly half an apsidal cycle earlier, the currently shallower eclipse was actually the deeper eclipse. In the case of binary B, the tendency is just the opposite, as can be seen in Fig.~\ref{fig:ABfolds}.\footnote{In this regard, we note that at the beginning of our comprehensive analysis we considered the deeper \textit{TESS} eclipses as primary ones in both binaries and the stars were labeled accordingly. Then, after concluding that the eclipse depths in both binaries reverse throughout an apsidal cycle and, moreover, considering the fact that currently, in the case of binary A the cooler star is eclipsed during the deeper light minima, we decided to redefine primary eclipses as those events in which the hotter (and more massive) stars are eclipsed, irrespective of the instantaneous amplitude ratio of the two eclipses. Hence, finally we relabeled the components of binary A, and all the tabulated parameters (and the figures) are given accordingly.}

 % Fig. 7

Turning to the apsidal motion cycles mentioned above, our direct spectral fits give a $\sim25$\,yr-period for both inner binaries. Allowing for the fact that the duration of the full spectroscopic dataset is $\sim2.7$\,yr, i.e., about 10\% of the full cycles, we feel that these results are in essentially perfect agreement with the findings of the spectro-photodynamical analysis (${P_\mathrm{apse}}_\mathrm{A,B}=28.7\pm0.1$ and $25.1\pm0.1$\,yrs, respectively). The latter results are based on the ETV data, which extend to more than the half of the full apsidal cycles. In addition to the apsidal motion periods of the inner binaries, the photodynamical solution also gives the apsidal motion for the outer orbit, and this is also found to be remarkably rapid at ${P_\mathrm{apse}}_\mathrm{AB}=145.3\pm0.2$\,yr. These values, which were `measured' from the numerical integration of the best-fit photodynamical solution, are in good accord with the theoretically calculated apsidal motion periods---of which the medians were found to be $P_\mathrm{apse}=31.8\pm0.1$,  $29.0\pm0.1$ and,  $151.6\pm0.8$\,yrs for orbits A, B, and AB, respectively.\footnote{The discrepancies of $3-6$\,yrs (i.e. 4-10\%) might have come from the neglect of the octupole order perturbation terms.} Besides the `measured' and theoretically calculated apsidal motion periods, we also tabulate the contribution of the dynamical (third-body), relativistic, and classic tidal effects to the apsidal advance rates ($\Delta\omega_\mathrm{3b, GR, tide}$, respectively). It is evident that the dynamical effects substantially dominate over the tidal and relativistic ones, and this clearly supports our previous statement in Sect.~\ref{sec:ETV_study} that the rapid apsidal motion must have a dynamical origin.

This latter statement leads us back to the previously mentioned contradiction between our results and those of \citet{2021ARep...65..826V} which will be discussed below.

\subsection{Comparison with the results of \citet{2021ARep...65..826V}}
%\subsection{Comparison with the results of Volkov et al. (2021)}

As mentioned in the Introduction, \citet{2021ARep...65..826V} also published a detailed analysis on this quadruple system. Their conclusions are remarkably different from our findings. Here, we discuss the origins of these discrepancies and make attempts to resolve them.

First, the most substantial difference is that they report an outer period of about 6.6\,yr ($2420\pm20$\,days), instead of the much shorter outer period of 121\,days that we report in this work. They arrived at this conclusion via an analysis of the ETV curves, where besides the evident large-amplitude sinusoidal variations characteristic of dynamically driven apsidal motion with a $\sim$25\,yr period, they detected additional cyclic variations with a period of $\sim$2400 days. They interpret the latter variations as the light-travel-time effect (LTTE) originating from the orbit of the two EBs around their common center of mass. Moreover, they also detect another variation of small amplitude in the high-precision \textit{TESS} Sector 7 eclipse times, which they claim might be due to a small $\sim60$-day-period libration (which they call `nutation') in the lines of the apsides of both binaries, caused by the other pair. They do not investigate, however, how such a relatively wide outer orbital separation would be able to produce such a short-period effect, nor how it could account for the rapid apsidal motion.

In contrast to their results, we found that these small-amplitude, so-called `nutations' are the principal third-body effects. These variations are, in fact, the same thing as we have seen in the ETVs of several tight, hierarchical triple stellar systems where the ETVs are dominated by the $P_\mathrm{out}$ timescale third-body perturbations due to the tertiary (which, in the present case, is also a binary itself).  These variations can be modeled both analytically \citep[see, e.~g.][]{borkovitsetal15,borkovitsetal16} and numerically, as, in the present work. Our photodynamical analysis shows clearly that these structures in the ETVs of both EB pairs can be explained with the mutual $P_\mathrm{out}$ timescale third-body perturbations due to the other EB, and consequently, the two binaries orbit around each other with a period of $P_\mathrm{out}=121.50\pm0.02$\,days (see the lower panels of Fig.~\ref{fig:etv}).  This conclusion makes this 2+2 quadruple momentarily the one with the shortest known outer period.

Regarding the longest period---the largest amplitude cycles on the ETVs (of $\sim$25 years) of both binaries---we agree with \citet{2021ARep...65..826V} that these arise from apsidal motion. We note, however, that this relatively short period of the apsidal motions in both binaries can be explained neither by combinations of the classic tidal effects and the general relativistic contribution in the EBs, nor by the gravitational perturbations of a relatively distant companion in an $\approx6$\,yr-period. The characteristic timescale of dynamically driven apsidal motion is proportional to $P_\mathrm{apse}\propto P_\mathrm{out}^2/P_\mathrm{in}$ which, in the case of $P_\mathrm{out}=2420$\,days yields $\approx4923$ and $\approx5457$\,yrs for binaries A and B, respectively. In contrast to this, as one can see, e.g., in the upper panels of Fig.~\ref{fig:etv}, and also in Table~\ref{tbl:simlightcurve}, where we tabulated the tidal, relativistic and third-body contributions to the apsidal advance rate separately, our photodynamical solution indeed produces the observed apsidal motions.

The most pressing issue regarding the ETV curves is the origin of the extra cyclic variations with a period of a few years. As was mentioned above, \citet{2021ARep...65..826V} find their period to be $\sim2420\pm40$\,d, while our complex photodynamical analysis (and the preliminary analytic ETV studies, as well) have resulted in a period close to half  the value they found. Below, we attempt to explain the origin of these `extra' ETV variations that occur on a timescale of $\sim$1200 days.

\subsection{Origin of the ETV variations on a timescale between $P_{\rm out}$ and $P_{\rm apse-node}$}

As mentioned above, the most interesting question about BU CMi is the origin of the extra, $\approx3$ or 6\,yr-period cyclic variations of the ETVs. This effect, at first sight, looks quite mysterious because these cycles appeared naturally in the numerically-generated ETVs during our modelling processes (see, e.g., the upper panels of Fig.~\ref{fig:etv}).  However, they do not appear even in our most detailed analytical ETV model, described in \citet{borkovitsetal15}. In other words, they are not present in our formulae based on the once-averaged and doubly-averaged octupole-order analytic perturbation theories of hierarchical triple stellar systems.\footnote{In a hierarchical triple system, the perturbations occur on three different, well separable time-scales. (i) The short-period periodic perturbations have a characteristic period proportional to the period $P_\mathrm{in}$ of the inner binary, and the relative amplitudes are related to $(P_\mathrm{in}/P_\mathrm{out})^2$; (ii) the medium-period periodic perturbations act on the time-scale of the outer orbit $P_\mathrm{out}$, while their amplitudes are proportional to $P_\mathrm{in}/P_\mathrm{out}$; and, finally (iii) the long-period perturbations (sometimes called `apse-node' perturbations) have a characteristic time-scale related to $P_\mathrm{out}^2/P_\mathrm{in}$, while their relative amplitudes are in the order of unity, i.e., the given orbital element might take any of its physically realistic values during that interval. Considering the analytic description of these perturbations, e.g., with a perturbed Hamiltonian, the three groups of perturbations are connected to those trigonometric terms in which the arguments contain: (i) the mean anomaly of the inner orbit; (ii) the mean anomaly of the outer orbit, but not that of the inner orbit; and, (iii) neither of the two mean anomalies.  Hence, averaging out both mean anomalies (double averaging) from the Hamiltonian, one can model analytically the long-term (and secular) perturbations. By contrast, when averaging out only the inner mean anomaly, the medium-period perturbations can be studied.}

\begin{figure*}
    \centering
    \includegraphics[width=0.48\linewidth]{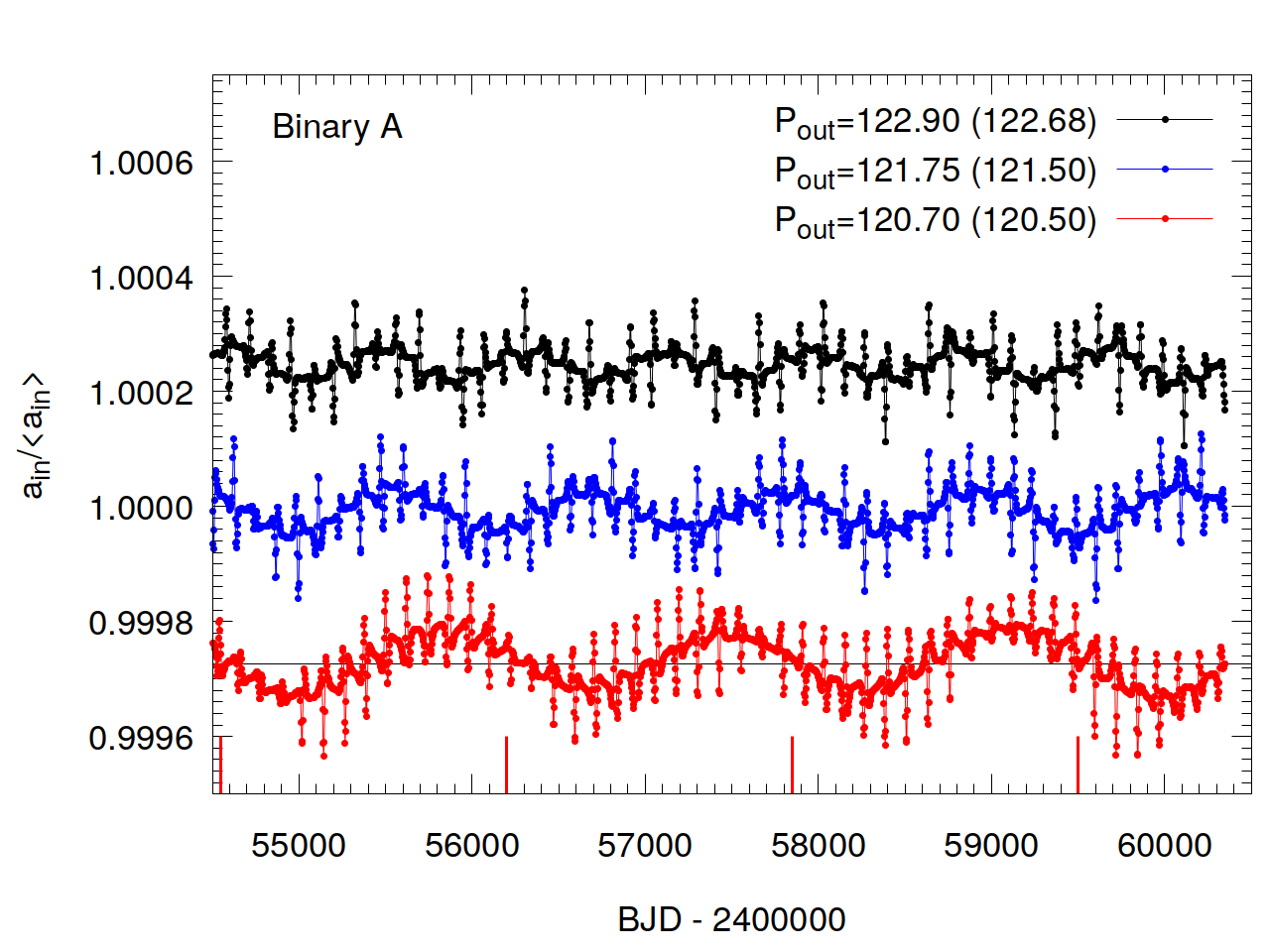}\includegraphics[width=0.48\linewidth]{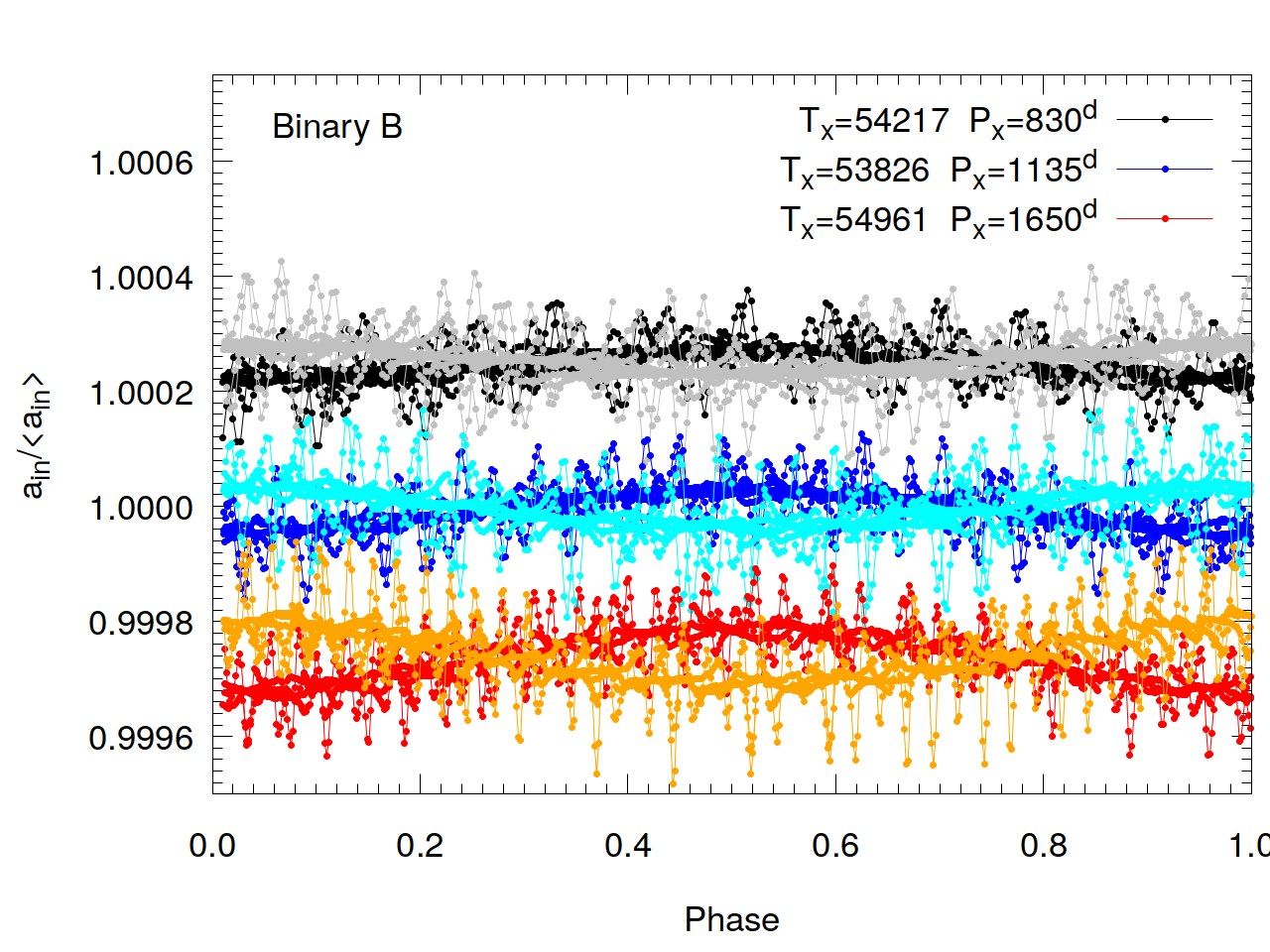}
    \includegraphics[width=0.48\linewidth]{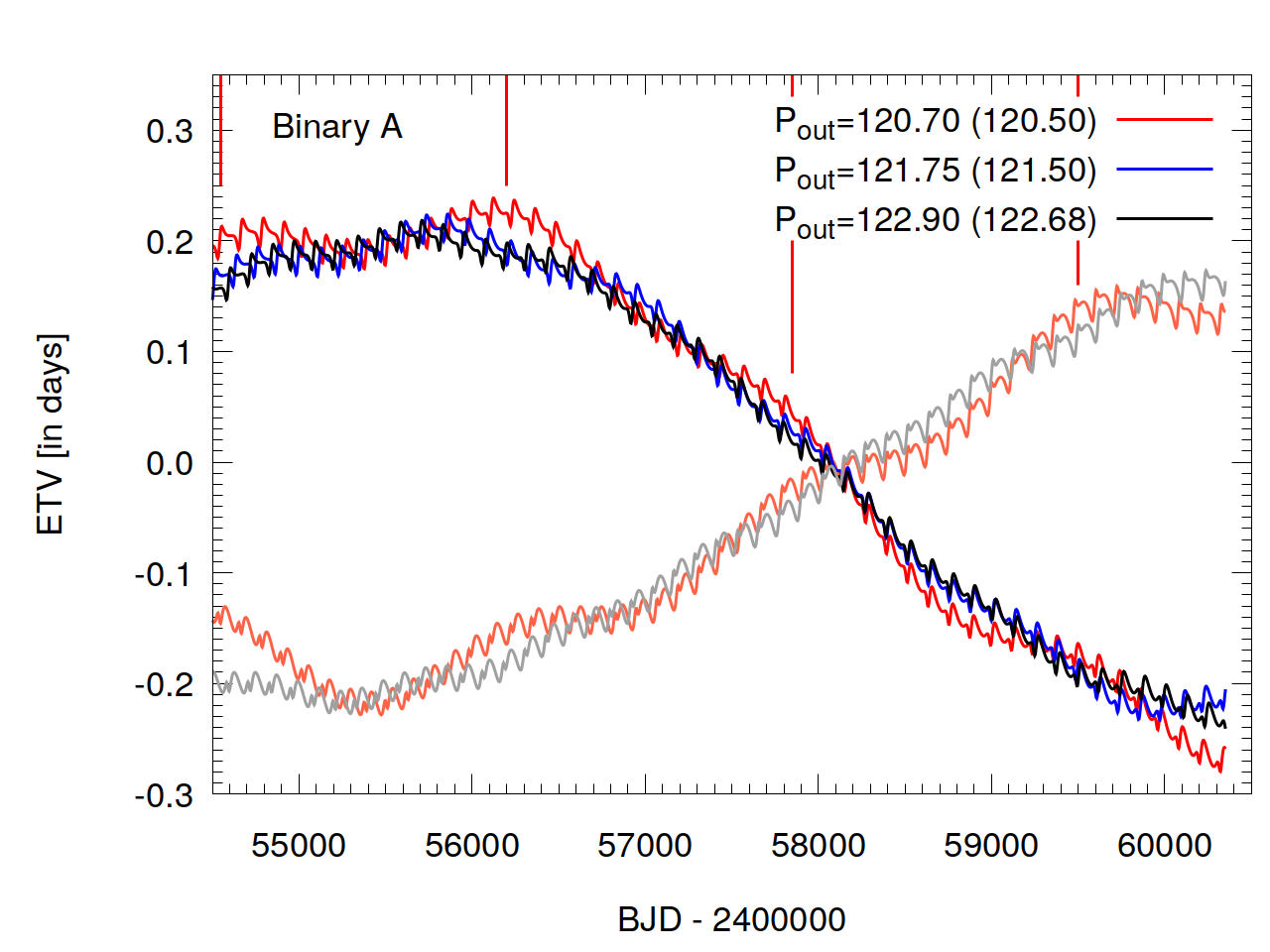}\includegraphics[width=0.48\linewidth]{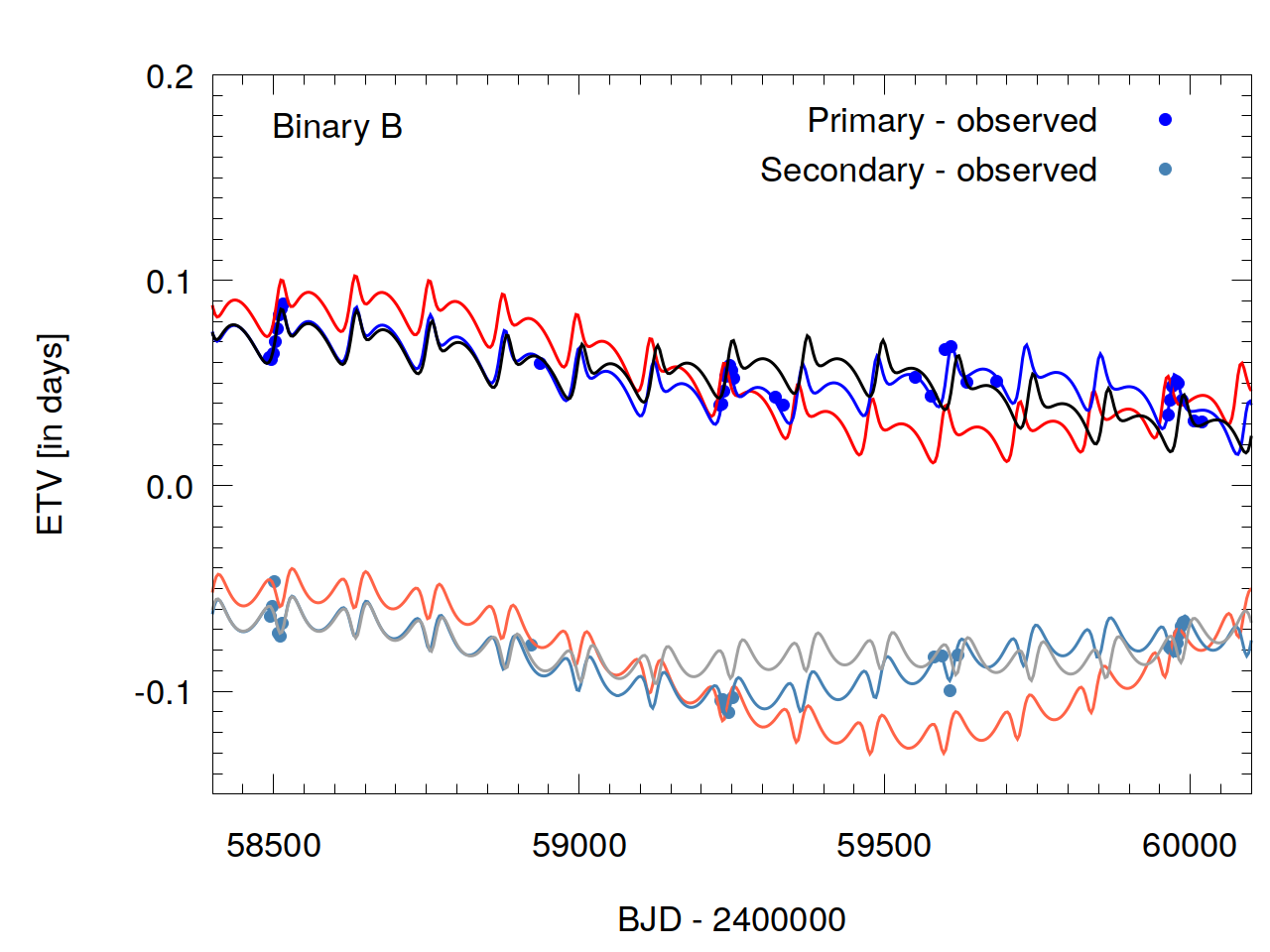}
    \caption{Upper left: The relative variations of the instantaneous osculating semi-major axes of binary A during three numerical integrations of the quadruple system. All the initial parameters of the integrations were set equal to the values of the best-fit solution, the only exception being the outer orbital period. (Note, in the legend, besides the osculating anomalistic outer periods which were used an input parameters for the given runs, we add also, in parentheses, the average outer periods derived from the results of the numerical integrations.) The semi-major axes were sampled at each integration step, but here we display only their averaged values over each inner (EB) period. {\it Upper right:} The variations of the semi-major axes of both binaries during the same three runs, but folded with the deduced periods of the variations. Semi-major axes of binary A are given with the same colors as in the left panel, while the corresponding quantities of binary B (folded with identical parameters) are plotted with lighter colors. The periods $P_\mathrm{x}$ used for the foldings are given in the legend. {\it Lower left:} Simulated ETV curves of binary A for the same interval than above. Darker colors represent primary ETV curves while the lighter ones are the secondary ones. (Note, from the secondary curves we left out the $P_\mathrm{out}=121.75$\,d-one for a better view.)  Vertical red lines in the left panels connect the positive extrema of the larger amplitude ETV bumps (the $P_\mathrm{out}=120.70$\,d case) with the corresponding (red) relative variations of the semi-major axis. The similar period and the phase shift of {90\degr} is evident. {\it Lower right:} Recent ETV curves of binary B according to the three currently investigated models vs. the observed times of minima. At this panel, calculating the ETV curves we used different periods for primary and secondary ETVs in order to get approximately horizontal lines. (With other words, the ETVs were calculated with the currently valid average primary and secondary periods.) As one can see, the observed ETV is in nice accord with the $P_\mathrm{out}=121.75$\,d-photodynamical model. Neither the $P_\mathrm{out}=120.70$\,d-model (with larger extra ETV bumps) nor the $P_\mathrm{out}=122.90$\,d-model (almost smooth ETV only with very low amplitude extra ETV bumps) cover the observations well.}
   \label{fig:axvar}
\end{figure*}

In order to identify the origin of these effects, we checked the variations of the instantaneous osculating orbital elements during each numerical integration step. In Fig.~\ref{fig:axvar} we show that we found periodic variations in the semi-major axes of the two binaries' orbits that are similar to what we see in the observed ETV curves. We display in the upper right panel of Fig.~\ref{fig:axvar}, that these variations have exactly the same periods and opposite phases in the two binaries, and hence, their effect on the ETVs may really mimic the signs of mutual LTTE. The connection between these small variations of the semi-major axes and the ETVs are also well illustrated in the lower left panel of Fig.~\ref{fig:axvar}. We can easily infer this connection not only for the similar periods, but also for the fact that the larger the amplitude of the variations in the semi-major axes, the larger the amplitude of the extra ETV bumps. And, moreover, one can see, that there is an exact 90\degr-phase shift in between the extrema of the semi-major axis and ETV variations. This is in perfect accordance with the fact that the ETV represent some cumulative or, integrated variations. Or, mathematically, a sine-like perturbation in the semi-major axis (or, in the mean motion) will result in a cosine-like ETV for the additional integration. This finding is not surprising, insofar as these slight variations in the semi-major axes naturally reflect the instantaneous mean motion of the given body, to which the variations of the ETVs are extremely sensitives. On the other hand, such kinds of variations in the semi-major axes are surprising in the sense that both the once- and doubly-averaged perturbation theories of hierarchical triple systems arrive at the conclusions that there are no medium-, long-period and secular perturbations in the semi-major axes. In other words, according to the perturbation theories that are usually used in the hierarchical stellar three-body problem, one should not expect any variations in the semi-major axes for which the periods exceed the period of the inner binary. Note, however, that in our last theoretical work about the analytical description of the ETVs of tight triples \citep{borkovitsetal15}, we introduced some further terms, which allow periodic perturbations in the inner semi-major axes on the timescale of the outer period; here, however, we detected periodic perturbations in these elements with a factor of $\sim$10 longer period.

In order to test the physical origin of this behaviour, we made some additional numerical runs where all but one of the initial parameters were the same (i.e., they were taken from the best-fit model), and only the outer period of the quadruple was slightly modified. As one can see in the upper panels of Fig.~\ref{fig:axvar}, only a small change in the outer period can result in substantial variations in the amplitude and the period of the effect, and this can also be seen clearly in the corresponding simulated ETVs (bottom panels of Fig.~\ref{fig:axvar}).

In generating and studying Fig.~\ref{fig:axvar}, we found it interesting to note that the closer the ratio of the outer to inner periods is to an integer, the larger the amplitude and longer the period are of these few-year-long cycles. For example, in the order of decreasing amplitudes, in binary A the period ratios are 40.99; 41.33; 41.73; while for binary B: 36.94; 37.24; 37.61.\footnote{For calculating these ratios, we did not use the instantaneous osculating anomalistic periods at epoch $t_0$ used as input parameters to the integrations; rather, from the results of the numerical integrations, we derived average outer periods, which are also given in the legends of the upper left panel of Fig.~\ref{fig:axvar}.} In our view, these facts suggest some similarities with the mean-motion resonances of the classical planetary perturbation theories.  Namely, when the mean motions are nearly commensurable this may lead to large-amplitude and long-period perturbations in the given orbital elements. Normally, however, only low-order mean motion resonances will produce large amplitude perturbations, because the amplitudes of the resonant perturbations in general are multiplied by some power-law functions of the eccentricities, in which the powers are proportional to the order of the resonances. Here, we suspect some similar behavior: Namely, due to the commensurability of the inner and outer periods, some very high order `resonant' terms yield some contribution which cannot be averaged out perfectly. This actually results in a very low-amplitude cyclic variation in the semi-major axes (the relative variations are of the order of $10^{-4} -10^{-5}$).  But, due to the extreme sensitivity of the ETVs to the mean-motions, these tiny variations may produce observable effects in the occurrence times of the eclipses.

We note that this effect is worthy of a more detailed and quantitative investigation. But it should be carried out first for a simpler configuration, i.e., for an actual 2+1 triple system instead of a 2+2 quadruple. In this regard we note that the photodynamical analysis of the tight triple TIC 167692429 \citep{borkovitsetal20}, which was based on the first year of \textit{TESS} data revealed similar extra cycles in the ETVs; these have now been verified by the third and fifth years of new \textit{TESS} observations. We will further investigate these extra cycles in the context of this triple system in the near future.

\section{Conclusions}
\label{sec:conclusions}

The bright variable star BU~CMi is composed of two short-period eclipsing binaries. Our work has found it to be the tightest quadruple system known, with $P_{\rm out} \simeq 121.5$ days.  Although the quadruple nature of the system was first pointed out by \citet{2021ARep...65..826V} from the \textit{MASCARA} photometry, they found a much longer outer orbital period of $P$ = 6.62 years, attributing the 121-day ETV variations to a `nutation' effect. Simultaneous modelling of high-dispersion spectroscopy and \textit{TESS} light curves conclusively demonstrates that the line profile variations cannot be explained by the 6.62-year periodicity. The spectroscopic data clearly show large, $\sim$50 km\,s$^{-1}$, variations in the systemic velocity of the inner subsystems (i.e., the two EBs). For $P_{\rm out}$ = 6.62 years, the amplitude of the RV changes of the binary centers of mass would be only about $\sim$ 18 km\,s$^{-1}$.

Even more conclusive proof of the extremely tight outer orbit comes from the detailed photodynamical modelling, which takes into account satellite photometry, all available eclipse timing data, and the RVs.  The numerical integration of the orbits explains the ETV changes with the 121-day period (the outer orbit), establishes the longer-term ETV variability on the $\sim$1200-day period, and perfectly predicts the rapid apsidal motions in the EBs on $\sim$25-year timescales. This apsidal motion is very rapid, and it is dynamically driven by the mutual gravitational perturbation of the binaries. For the much longer outer orbit determined by \citet{2021ARep...65..826V}, the apsidal motion rate would be substantially slower.

Although the orbital and stellar component parameters are well constrained by the complex photodynamical modeling, there remain a few open questions that require further observations.  For example, we seek to constrain any spin-orbit misalignment; this possibility, however, appears unlikely due to the near-coplanarity of the inner and outer orbits. Determination of the spin-orbit misalignment would require dedicated spectroscopy of the system during the eclipses of the stellar components. The small observed range of the radial velocities of the components and their high rotation rates would complicate the modelling of the Rossiter-McLaughlin effect.

Assuming that the stellar rotational rates are quasi-synchronous (rotation rate equals the Keplerian orbital rate at periastron) and that the spin axes are perpendicular to the inner orbits, we can further constrain the stellar radii from their measured projected rotational velocities $v \sin i$.

Although both of the inner binaries are eclipsing, the relatively low inclination of the outer orbit, $i_{A-B} \sim 83.8 \degr$, and a relatively large semimajor axis, $a_{A-B} \sim 221$ R$_\odot$, preclude outer-orbit eclipses. A shallow outer eclipse could occur if the outer inclination angle were larger than about 88.5\degr. 

Finally, we note that the BU CMi system could have had an outer period as short as $\approx 32$ days and still be dynamically stable (see, e.g., \citealt{mardling01}; \citealt{mikkola08}; \citealt{rappaport13}).  This assumes the same EB periods and masses, the same outer eccentricity, and the same orbital coplanarity. Thus, there is much room in phase space to find even tighter quadruples.  Whether evolution scenarios will permit such short-period quadruples is another matter.  It is therefore worth trying to observationally push the boundaries to ever shorter outer periods.

\section{Data availability}

All photometric and spectroscopic data used in this paper and the codes used for the direct fitting of the spectra and photodynamical analysis will be shared upon a reasonable request to the corresponding author.

\section*{Acknowledgements}

A.\,P. acknowledges the financial support of the Hungarian National Research, Development and Innovation Office -- NKFIH Grant K-138962.  TP and RK acknowledge support from the Slovak Research and Development Agency – contract No. APVV-20-0148 and the VEGA grant of the Slovak Academy of Sciences No. 2/0031/22. GB, ZC and JH acknowledge funding from NASA Grant 80NSSC22K0315. We would also like to thank the Pierre Auger Collaboration for the use of its facilities. The operation of the robotic telescope FRAM is supported by the grant of the Ministry of Education of the Czech Republic LM2023032. The data calibration and analysis related to the FRAM telescope is supported by the Ministry of Education of the Czech Republic MSMT-CR LTT18004, MSMT/EU funds CZ.02.1.01/0.0/0.0/16$\_$013/0001402, CZ.02.1.01/0.0/0.0/18$\_$046/0016010 and CZ.02.1.01/0.0/0.0/18$\_$046/0016007.

This paper includes data collected by the \textit{TESS} mission. Funding for \textit{TESS} is provided by NASA's Science Mission Directorate.

\bibliography{bu_cmi}{}
\bibliographystyle{mnras}



\label{lastpage}
\end{document}